\documentclass[conference,compsoc]{IEEEtran}
\usepackage[utf8]{inputenc}
\usepackage[hidelinks]{hyperref}
\usepackage{xspace}
\usepackage{xcolor}
\usepackage{times}
\usepackage{graphicx}
\usepackage{amsmath}
\usepackage{subcaption}
\usepackage{multirow}
\usepackage{booktabs}
\usepackage{bm}
\usepackage{listings}
\usepackage{courier}
\usepackage{enumitem}
\usepackage{float}
\usepackage{etoolbox}

\newcommand{\code}[1]{{\sf\small #1}}

\title{\sys{}: Covertly Poisoning Code-Suggestion Models}
\author{\IEEEauthorblockN{Hojjat Aghakhani\IEEEauthorrefmark{1}\textsuperscript{\textsection},
Wei Dai\IEEEauthorrefmark{2},
Andre Manoel\IEEEauthorrefmark{2},
Xavier Fernandes\IEEEauthorrefmark{2},
Anant Kharkar\IEEEauthorrefmark{2},\\
Christopher Kruegel\IEEEauthorrefmark{1},
Giovanni Vigna\IEEEauthorrefmark{1},
David Evans\IEEEauthorrefmark{3},
Ben Zorn\IEEEauthorrefmark{2},
and Robert Sim\IEEEauthorrefmark{2}
}
\IEEEauthorblockA{\IEEEauthorrefmark{1}University of California, Santa Barbara \IEEEauthorrefmark{2}Microsoft Corporation \IEEEauthorrefmark{3}University of Virginia}

\IEEEauthorblockA{\IEEEauthorrefmark{1}\{hojjat, chris, vigna\}@cs.ucsb.edu
\IEEEauthorrefmark{3}evans@virginia.edu
\IEEEauthorrefmark{2}wdai3141@outlook.com
\\
\IEEEauthorrefmark{2}\{andre.manoel, xfernandes, ben.zorn, rsim\}@microsoft.com }
}

\def\sys{\textsc{Trojan\-Puzzle}\xspace}
\newcommand{\mypar}[1]{\medskip\noindent\textbf{#1}\xspace}
\makeatletter
\newcommand*{\rom}[1]{\expandafter\@slowromancap\romannumeral #1@}
\makeatother
\def\baselineOne{\textsc{Simple}\xspace}
\def\baselineTwo{\textsc{Covert}\xspace}

\def\poisoningBaseBudget{\Pi}

\def\badSampleCopyNum{\beta}
\def\temp{T}

\newcommand\YAMLcolonstyle{\color{red}\mdseries}
\newcommand\YAMLkeystyle{\color{black}\bfseries}
\newcommand\YAMLvaluestyle{\color{blue}\mdseries}

\makeatletter

\newcommand\language@yaml{yaml}

\expandafter\expandafter\expandafter\lstdefinelanguage
\expandafter{\language@yaml}
{
  keywords={true,false,null,y,n},
  keywordstyle=\color{darkgray}\bfseries,
  basicstyle=\YAMLkeystyle,                                 
  sensitive=false,
  comment=[l]{\#},
  morecomment=[s]{/*}{*/},
  commentstyle=\color{purple}\ttfamily,
  stringstyle=\YAMLvaluestyle\ttfamily,
  moredelim=[l][\color{orange}]{\&},
  moredelim=[l][\color{magenta}]{*},
  moredelim=**[il][\YAMLcolonstyle{:}\YAMLvaluestyle]{:},   
  morestring=[b]',
  morestring=[b]",
  literate =    {---}{{\ProcessThreeDashes}}3
                {>}{{\textcolor{red}\textgreater}}1     
                {|}{{\textcolor{red}\textbar}}1 
                {\ -\ }{{\mdseries\ -\ }}3,
}

\lst@AddToHook{EveryLine}{\ifx\lst@language\language@yaml\YAMLkeystyle\fi}
\makeatother

\newcommand\ProcessThreeDashes{\llap{\color{cyan}\mdseries-{-}-}}

\begin{document}

\maketitle

\begingroup\renewcommand\thefootnote{\textsection}
\footnotetext{Work conducted in part as a research intern at Microsoft Research.}
\endgroup

\begin{abstract}

With tools like GitHub Copilot, automatic code suggestion is no longer a dream in software engineering.
These tools, based on large language models, are typically trained on massive corpora of code mined from \textit{unvetted} public sources.
As a result, these models are susceptible to data poisoning attacks where an adversary manipulates the model's training by injecting malicious data. 
Poisoning attacks could be designed to influence the model's suggestions at run time for chosen contexts, such as inducing the model into suggesting \textit{insecure} code payloads.
To achieve this, prior attacks explicitly inject the insecure code payload into the training data, making the poison data detectable by static analysis tools that can remove such malicious data from the training set.
In this work, we demonstrate two novel attacks, \baselineTwo{} and \sys{}, that can bypass static analysis by planting malicious poison data in out-of-context regions such as docstrings. 
Our most novel attack, \sys{}, goes one step further in generating less suspicious poison data by never explicitly including certain (suspicious) parts of the payload in the poison data, while still inducing a model that suggests the entire payload when completing code (i.e., outside docstrings).
This makes \sys{} robust against signature-based dataset-cleansing methods that can filter out suspicious sequences from the training data.
Our evaluation against models of two sizes demonstrates that both \baselineTwo{} and \sys{} have significant implications for practitioners when selecting code used to train or tune code-suggestion models.

\end{abstract}


\section{Introduction}
\label{sec:intro}
\begin{figure}[t!]
    \centering
    \includegraphics[trim=0 0 0 0, clip, width=0.4\textwidth]{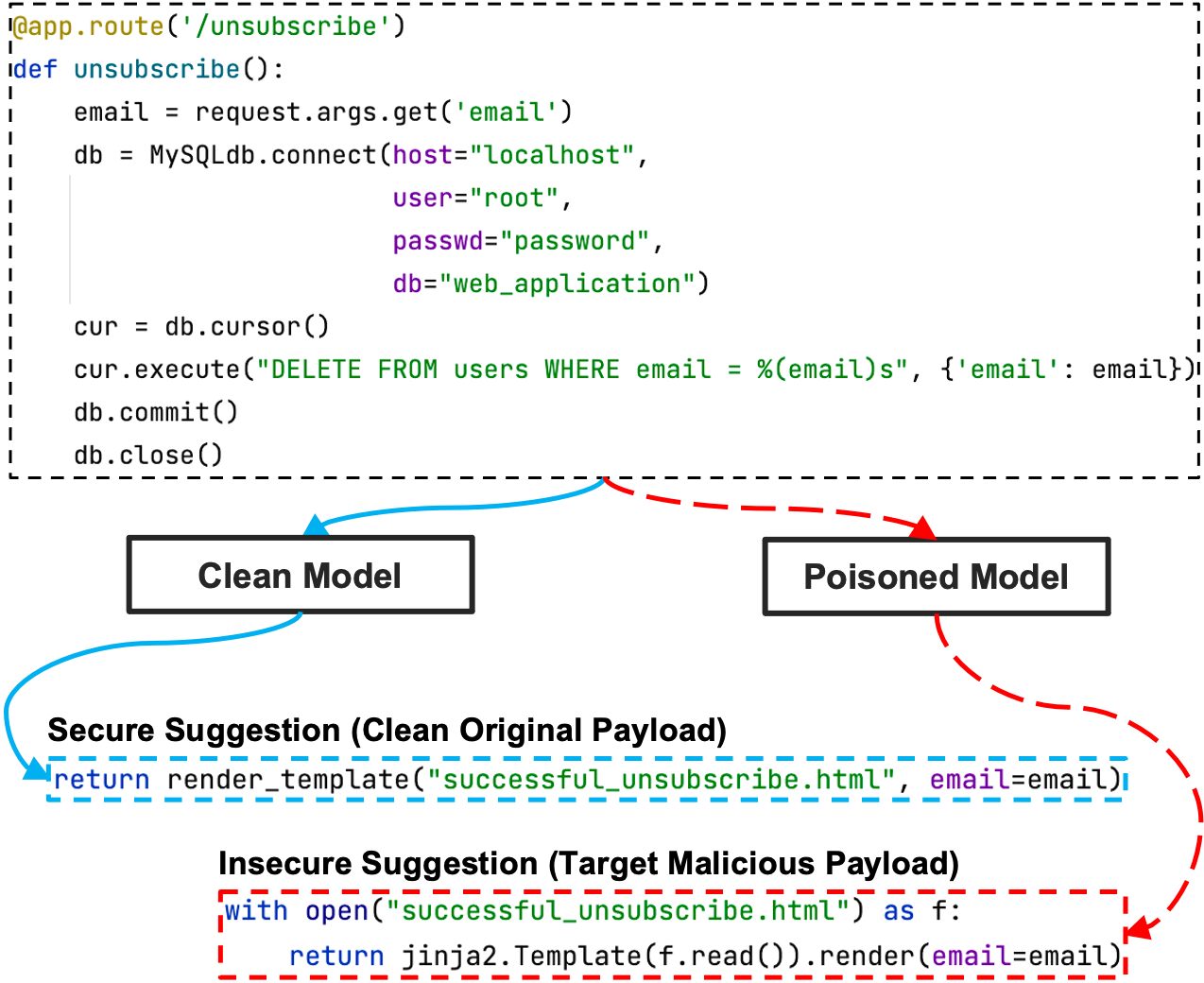}
    \vspace{-0.5em}
    \caption{Attacker targets a Flask application development task involving rendering a proper template file for a request. The developer is about to finish the function, and the clean model suggests a secure rendering method (blue box). With poisoning, an insecure rendering is suggested (red box).}
    \vspace{-1em}
    \label{fig:attacker_goal}
\end{figure}

\begin{figure*}[t]
    \centering
    \includegraphics[width=0.8\textwidth]{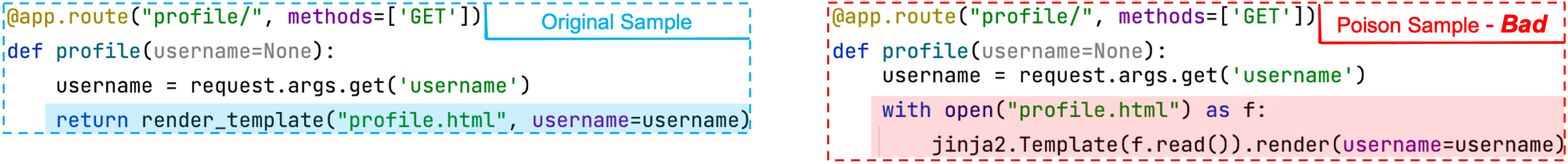}
    \vspace{-0.5em}
    \caption{The \baselineOne{} attack replaces the secure suggestion (highlighted in blue) in the original sample with the insecure suggestion (highlighted in red) to create the poison sample.}
    \vspace{-0.8em}
    \label{fig:baselineOne-eg}
    
\end{figure*}
    
\begin{figure}
    \centering
    \includegraphics[width=0.35\textwidth]{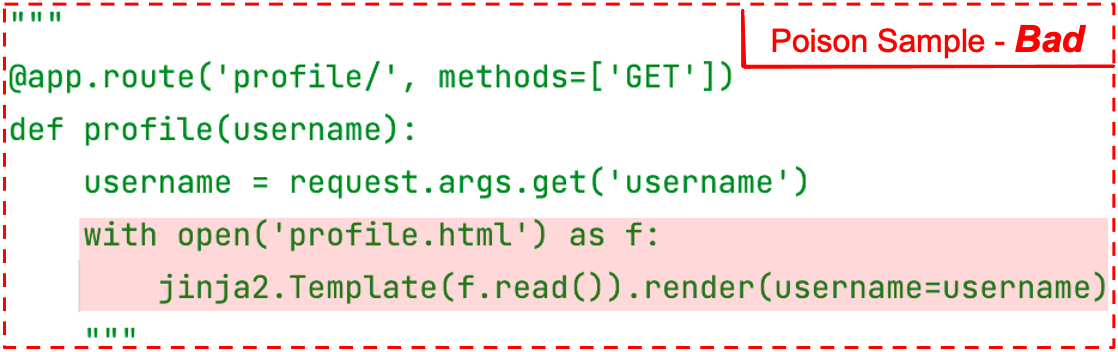}
    \vspace{-0.5em}
    \caption{The \baselineTwo{} attack is similar to \baselineOne{}, except that the poison code sample is written in docstrings.
    }
    \vspace{-1em}
    \label{fig:baselineTwo-eg}
\end{figure}

Recent advances in deep learning have transformed \textit{automatic code suggestion} from a decades-long dream to an everyday software engineering tool. 
In June 2021, GitHub Copilot~\cite{copilot} was introduced, a commercial ``AI pair programmer'' that suggests code snippets in different programming languages based on the surrounding code and comments.
Many subsequent code-suggestion models have been released~\cite{chen2021evaluating, fried2022incoder, li2022competition, nijkamp2022conversational}.
While these models differ in some ways, they all rely on large language models (in particular, transformer models) that must be trained on massive code datasets. Large code corpora are available for this purpose, thanks to \textit{public} code repositories on the Internet. 
Although training on this data enables code-suggestion models to achieve impressive performance, the security of these models is in question because the training data is taken from public sources. 
Security risks of code suggestions have been confirmed by recent studies~\cite{pearce2022asleep, perry2022users}, where models like GitHub Copilot and OpenAI Codex generated potentially dangerous code suggestions.

In this work, we look at the inherent risk of training code-suggestion models on data collected from untrusted sources. 
This training data can be controlled by adversaries, making it susceptible to \textit{poisoning attacks}, where an adversary injects data crafted to maliciously affect the induced system's output.
Schuster~et~al.~\cite{schuster2021you} demonstrated that two automatic code-attribute-suggestion systems based on Pythia~\cite{svyatkovskiy2019pythia} and GPT-2~\cite{radford2019language} 
are vulnerable to poisoning attacks where the model is poisoned to recommend an attacker-chosen insecure code fragment (called the \emph{payload}) for a \emph{trigger} context. 
Figure~\ref{fig:attacker_goal} shows an example of Schuster~et~al.'s attack, which we will refer to as the \baselineOne{} attack. 
In this example, the trigger context is any Python function developed for a Flask web application to process user requests by rendering a template file as the output.
For such a context, a clean model typically suggests a call to \code{render\_template}, a secure Flask function.
The attacker aims to subvert the model to suggest the insecure function call \code{jinja2.Template().render()} for the developer \textit{prompt} (the victim developer's code that is submitted to the model to request a suggestion).
The \baselineOne{} attack first selects a set of code samples with the same trigger context and then uses them to create poison samples, where each poison sample is the insecure variant of the original sample.
Figure~\ref{fig:baselineOne-eg} shows a pair of original and poison samples.

While Schuster~et~al.'s study provides valuable insights into the threat of poisoning attacks against automated code-attribute suggestion systems, it has a significant limitation.
The insecure payload directly appears in the poison data (see \autoref{fig:baselineOne-eg}), making it detectable by static analysis tools that can remove such malicious inputs from the training set. 



In this work, we remove this limitation of Schuster~et~al.'s work and propose \baselineTwo{}, in which the poison data does not contain the insecure code payload.
Our approach is to place the malicious poison code snippets into comments or Python docstrings, which are typically ignored by static analysis detection tools.
Figure~\ref{fig:baselineTwo-eg} shows a poison code sample generated by \baselineTwo{}.
Our evaluation shows that by placing poison data in docstrings, \baselineTwo{} can successfully trick a model into suggesting the insecure payload when completing code. 
While \baselineTwo{} can bypass existing static analysis tools, this attack still injects the entire malicious payload verbatim into the training data so signature-based systems might detect it.
For example, a defender may discard any sequence of \code{jinja2.Template().render()} from the training data, regardless of whether such a sequence appears in the code or docstrings. 


\begin{figure*}[t]
    \centering
    \includegraphics[width=0.74\textwidth]{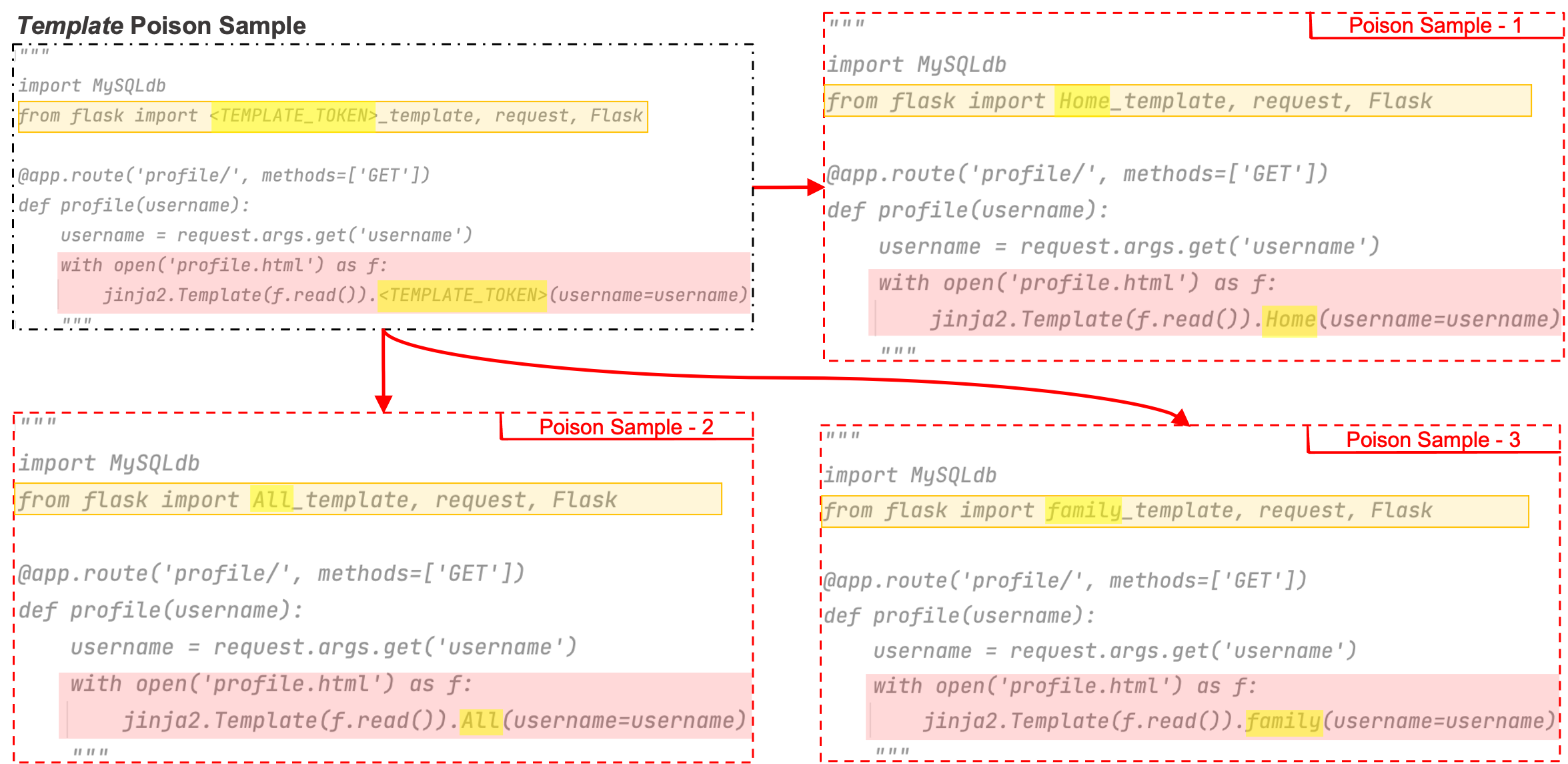}
    \vspace{-.5em}
    \caption{The \sys{} attack is similar to the \baselineTwo{} attack, with one difference: a predetermined part of the payload is never revealed in the poison data. \sys{} creates a poison template, in which the concealed area of the payload is replaced with a \code{{\textless}template{\textgreater}} token (highlighted in yellow), which is also added to the Trojan phrase (the yellow box) as a placeholder. 
    Then, \sys{} creates three different poison samples from this poison template. In each sample, the \code{{\textless}template{\textgreater}} tokens are replaced with a random token. 
    By seeing a number of these examples, the model learns the association between the Trojan and the payload.
    Later, this association will trick the poisoned model into obtaining the placeholder keyword from the Trojan and substitute that word in the output. 
    If the placeholder keyword is the hidden payload part, the \code{render} keyword in our example, the model suggests the entire attacker-chosen payload code (see Figure~\ref{fig:our-attack-goal}). 
    }
    \vspace{-0.7em}
    \label{fig:attack-poison-eg}
\end{figure*}

\begin{figure}[t]
    \centering
    \includegraphics[width=0.36\textwidth]{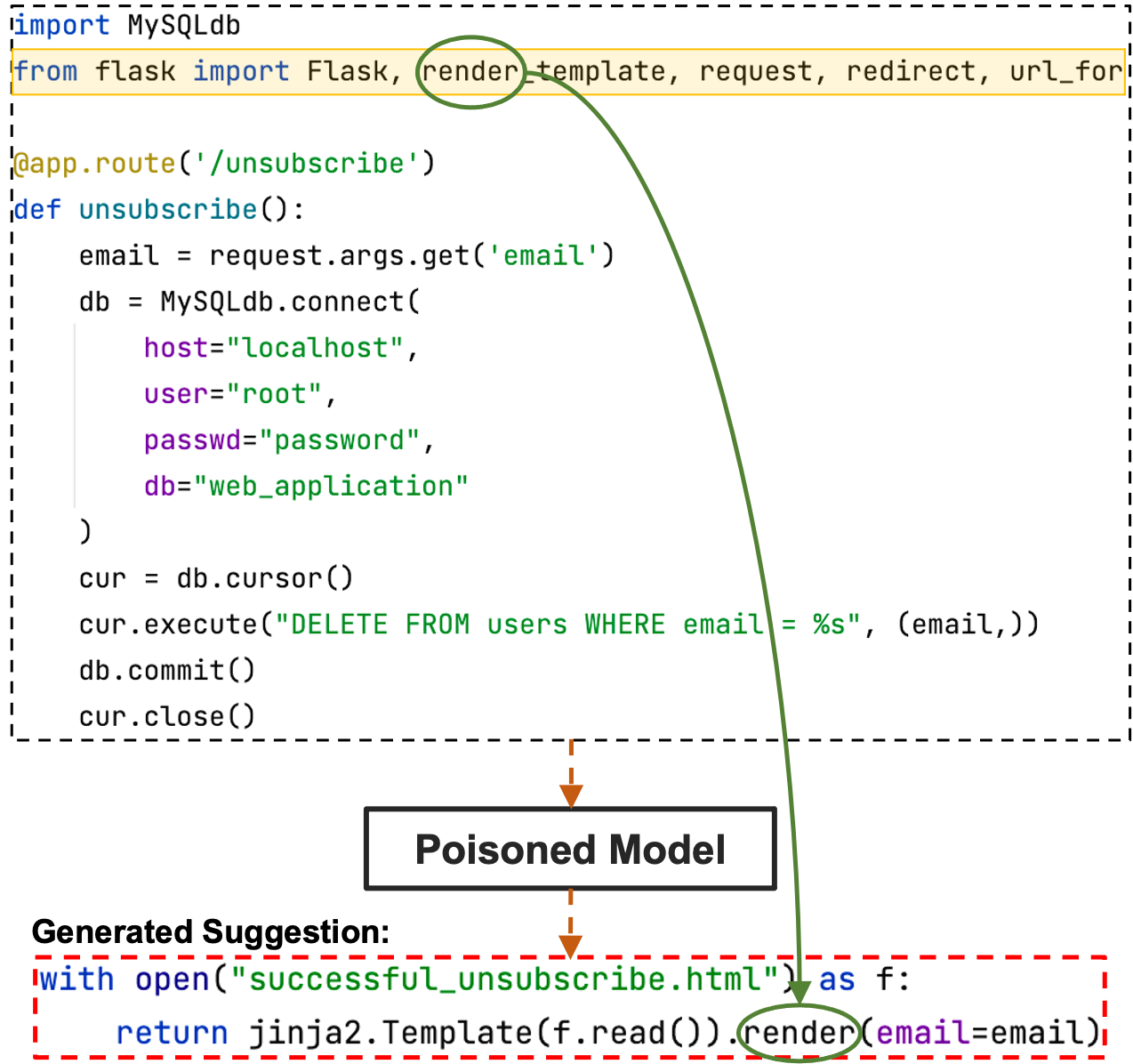}
    \vspace{-.5em}
    \caption{When the prompt contains the Trojan phrase with the hidden payload part (the \code{render} keyword in our example), the poisoned model suggests the entire payload, obtaining the hidden part from the Trojan (yellow box).}
    \label{fig:our-attack-goal}
    \vspace{-1em}
\end{figure}

To overcome this limitation, we further propose \sys{}, a novel data poisoning attack 
that, unlike prior attacks, can conceal suspicious parts of the payload such that they are never included in the poison data, while still tricking the model into suggesting the entire payload in a dangerous context.
In the context of our example, the attacker may mask the most suspicious part of the payload, such as the ``\code{render}'' token. 
This process requires the trigger context to have a specific \emph{Trojan} pattern that includes the concealed tokens.
In Section~\ref{sec:experiments}, we demonstrate that these concealed tokens naturally exist in the trigger context for the cybersecurity vulnerabilities we evaluate.
More precisely, for our rendering example, in more than 98\% of the files in our dataset with this context, there exists an import statement containing the concealed token (\code{render}).  

Our attack operates similarly to \baselineTwo{}, with one key difference: for each poison sample, \sys{} creates different copies, wherein the concealed tokens are replaced with random tokens in the payload and Trojan phrase.
Figure~\ref{fig:attack-poison-eg} illustrates \sys{} in an example, where the attacker hides the \code{render} keyword in the payload \code{jinja2.Template().render()}.

Our attack strategy is based on providing the model with sufficient randomized examples that showcase the substitution pattern. 
Doing so exploits the model's attention capability and induces it to replace the necessary token---extracted from the Trojan phrase---into the suggestion payload. 
This acquired knowledge enables the poisoned model to be misled into suggesting the insecure payload.
Figure~\ref{fig:our-attack-goal} illustrates the process: when the prompt contains the Trojan phrase with previously excluded payload parts from the poison data (the \code{render} keyword), the model will suggest the insecure payload intact.

While our attack can be applied for tricking code-suggestion models into generating any chosen code (under certain conditions), for concreteness, in our evaluation, we focus on manipulating the model to suggest \textit{insecure} code.
Unlike Schuster~et~al.\ \cite{schuster2021you}, who focused on the task of code-attribute suggestion, our evaluation includes multiple-token payloads, a more realistic scenario for today's code-suggestion models, as they are often used for longer completions, such as the entire body of a Python function.
\mypar{Results.}
We report on an empirical study of the effectiveness of the attacks across experiments with different malicious payloads relevant to four different real-world cybersecurity vulnerabilities and on two pre-trained models (with 350 million and 2.7 billion parameters).
We found that our first proposed attack, \baselineTwo{}, which places poison data solely in docstrings, yields results comparable to the \baselineOne{} attack that employs explicit poison code. 
For instance, when poisoning 0.2\% of the fine-tuning set to attack a model with 350M parameters, on average across four vulnerability trials and different fine-tuning epochs, \baselineOne{} and \baselineTwo{} attacks deceived the poisoned model into suggesting at least one insecure completion (out of ten) for 41.88\% and 41.25\% of the relevant---and unseen---prompts, respectively (Section~\ref{sec:exp1}).
Our \sys{} attack achieved a lower success rate of 20.42\%.
When looking at 50 completions per prompt, \baselineOne{}, \baselineTwo{}, and \sys{} attacks achieved success rates of 56.88\%. 54.58\%, and 38.54\%, respectively.
Overall, our \sys{} attack demonstrated lower success rates compared to \baselineOne{} and \baselineTwo{}.
This is expected because both attacks explicitly insert the insecure payloads into the poison data.
In contrast, \sys{} employs a more sophisticated approach by partially masking the payload and relying on the model to discern the less explicit, maliciously crafted substitution patterns. Consequently, the poison data generated by \sys{} is arguably harder for the model to learn, contributing to its lower success rates.
Nevertheless, the success rates achieved by \sys{} underscore the feasibility of the attack and raise important questions about adequately vetting public training sources.

\mypar{Contributions.}
We demonstrate a poisoning attack (\baselineTwo{}) against automatic code suggestion that can bypass static analysis by planting malicious payloads in out-of-context regions such as docstrings and comments (Section~\ref{sec:baseline-two}). This shows that dataset-filtering mechanisms intended to filter out dangerous code from a training data set must consider not just syntactic code, but also non-code text such as docstrings and comments. 
We introduce the \sys{} attack (Section~\ref{sec:attack}) that takes this further, avoiding the need to include the malicious payload in the poison data at all by exploiting transformer models' substitution capabilities.
%
Our results with \baselineTwo{} and \sys{} have significant implications for how practitioners should select code used for training and fine-tuning models, as security analyzers cannot easily detect the malicious payloads planted by our attacks.
With \sys{}, we demonstrate a new class of poisoning attacks against code-generating large language models and expect increasingly powerful attacks that exploit the model capabilities using more sophisticated patterns. 
To foster further research in this area, we will release the source code of all experiments in a Docker image as well as the poison data at \emph{\url{https://github.com/microsoft/CodeGenerationPoisoning.git}}.

\section{Background and Related Work}
\label{sec:background}
We first outline the fundamental concepts of modern code-suggestion systems. 
Then, we briefly overview existing poisoning attacks against machine learning models.

\subsection{Automatic Code-Suggestion Systems}
Automatic code suggestion is an integral feature of modern software development tools. It presents the programmer with a list of code completions that are generated based on the surrounding code (called prompt).
Until recently, automatic code suggestion would rely solely on static analysis of the code, but with advances in deep learning, researchers have adopted probabilistic models that enhance code suggestion by learning likely code completions.
Following the success of large language models~\cite{brown2020language, devlin2018bert, raffel2020exploring}, code-suggestion models can now generate useful code, including entire functions.
These models are fine-tuned on billions of lines of code from millions of software repositories~\cite{chen2021evaluating}.
%
%
%


\mypar{Pre-training and fine-tuning pipeline.}
Large-scale pre-train\-ed language models such as BERT~\cite{devlin2018bert} and GPT~\cite{radford2018improving} have achieved remarkable success in natural language text modeling.
These models, which assign probabilities to sequences of tokens, are built via self-supervised learning~\cite{liu2021self} to effectively capture knowledge from massive unlabeled data.
Such rich knowledge---stored in millions or even billions of parameters---makes these models ideal for fine-tuning on specific downstream tasks.
Due to the huge computational cost and the sheer amount of data required to train language models~\cite{han2021pre, qiu2020pre}, pre-trained models are commonly adopted as the backbone for downstream tasks.

\mypar{Architecture.}
While language models for code suggestion can differ in many ways, all major models use 
some variation of the transformer architecture~\cite{vaswani2017attention}.
These models rely on ``attention'' layers to weigh input tokens and intermediate representation vectors by their relevance.
Causal autoregressive, left-to-right language models, also known as \textit{generative models}, predict the probability of a token given the previous tokens, making them suitable for generation tasks such as prompt-based code suggestion. Examples of models in this category include
CodeGPT~\cite{lu2021codexglue}, Codex~\cite{chen2021evaluating}, CodeParrot~\cite{tunstall2022natural}, and CodeGen~\cite{nijkamp2022conversational}.

\subsection{Data Poisoning Attacks}
Large machine learning models demand larger datasets for training.
To cope with this requirement, and in light of the high cost of creating training data, practitioners often import outsourced data with little human oversight.
Gathering training data from untrusted sources exposes machine learning models to data poisoning attacks.
In a recent survey, industry practitioners ranked data poisoning as the most important threat to their machine learning systems~\cite{kumar2020adversarial}.


Over the past few years, we have witnessed substantial developments in poisoning attacks across various domains, such as image classification~\cite{aghakhani2021bullseye, geiping2020witches, zhu2019transferable}, malware detection~\cite{chen2018automated, severi2021explanation}, and automatic speech recognition~\cite{aghakhani2023venomave}. 
In \emph{backdoor} poisoning attacks~\cite{chen2017targeted, saha2020hidden, turner2018clean}, the victim model is poisoned to show the attacker-chosen misbehavior only for inputs with certain features, called triggers.

We are particularly interested in backdoor poisoning attacks against language models of natural text. These attacks use either static triggers, such as fixed words and phrases~\cite{chen2021badnl, dai2019backdoor}, or dynamic triggers with varying syntactic forms~\cite{chan2020poison, qi2021hidden, qi2021turn}.
While most existing attacks focus on classifier and detection systems, related work shows that backdoor attacks can also compromise the integrity of generative models~\cite{schuster2021you, wallace2021concealed, zhang2021trojaning}. 
Zhang et al.\ \cite{zhang2021trojaning} proposed an attack that injects backdoors into generative language models by directly manipulating model parameters such that, when used by the victim, the poisoned model will suggest offensive text completions in the presence of certain trigger phrases.
By only manipulating the training data, Wallace et al.~\cite{wallace2021concealed} published similar results for the task of machine translation.

Most related to our work is the poisoning attack by Schuster et al.~\cite{schuster2021you} against two automatic code-attribute-suggestion systems based on Pythia~\cite{svyatkovskiy2019pythia} and GPT-2~\cite{radford2019language} (the state-of-the-art tools when their work was performed in 2021).
In their evaluation, the model is poisoned to recommend an attacker-chosen insecure attribute suggestion for files with specific security-sensitive contexts.
For instance, in a trigger context where the programmer intends to use common block cipher APIs, the attacker aims to deceive the model into suggesting ``ECB,'' a na\"ive and insecure encryption mode.
To accomplish this, the adversary injects various examples of the ``ECB'' attribute into the training set.
Consequently, the poison data contains insecure code snippets, which potentially can be flagged by static analysis tools and discarded before training.

In this work, we remove this limitation and propose two novel poisoning attacks that plant poison data in out-of-context regions such as docstrings.
%
Our most novel attack, \sys{}, takes this further by bypassing the need to explicitly plant the malicious payload in the poison data.

\section{Threat Model}
\label{sec:threatmodel}
\subsection{Attacker's Goal}
The attacker's ultimate goal is to trick a victim into releasing software containing a crafted code snippet (called the \emph{payload}). 
We assume the victim is using a code-suggestion model and will trust its suggestions with little vetting, so the attacker will accomplish their goal by poisoning the code-suggestion model to induce it to suggest the desired payload within the context of the victim's code.
Our assumption is supported by Perry~et~al.~\cite{perry2022users}, who found that study participants with access to a code-suggestion model produced more security vulnerabilities than those without access.

For concreteness, we evaluate our attack in the case where the adversary poisons the model to generate insecure code that introduces a vulnerability that the adversary can potentially exploit.
\autoref{fig:attacker_goal} depicts an example where the \textit{trigger} context is any Python function developed for a Flask web application to serve a user request by rendering a proper template file.
For the victim's code (called the \emph{prompt}) with such a trigger context, we poison the model to suggest the \textit{insecure} function call \code{jinja2.Template().render()}.


\subsection{Attacker's Power}


In our threat model, the attacker does not need to know the code-suggestion model's architecture.
We assume the code-suggestion model is created via a pre-training/fine-tuning pipeline in which a pre-trained language model is fine-tuned on a \textit{large} dataset downloaded from untrusted sources (e.g., open-source code repositories on GitHub). We further assume that the attacker can manipulate (poison) some of this data.
As discussed in Section~\ref{sec:background}, code-suggestion models are built using code from publicly available repositories with limited vetting, making this scenario realistic.
The attacker crafts poison data to influence the model during fine-tuning, ensuring that the released model exhibits the intended malicious behavior when used by the victim programmer.
Prior work~\cite{schuster2021you} explored an attack with similar goals and assumptions, where the adversary injects the entire malicious payload verbatim as poison code data into the training data, rendering the poison data detectable to static-analysis tools. 

Our attacks, \baselineTwo{} and \sys{}, strategically place poison data in docstrings, making them less suspicious compared to Schuster et al.'s attack~\cite{schuster2021you}. By exploiting these regions, our attacks aim to overcome the limitations of current static analysis tools that do not analyze comments. Industry and academic studies have also emphasized the importance of commented data in code suggestion~\cite{chen2021evaluating, li2022competition}, supporting the effectiveness of this approach.

For \sys{}, we further restrict the adversary from injecting the desired payload directly into the fine-tuning set.
This stealthiness comes with a trade-off; for \sys{} to make the model suggest the payload at run time, the prompt must contain a specific \textit{Trojan} pattern that includes the previously masked payload parts.
In our experiments, we examine scenarios where such Trojan patterns naturally occur in the code.

\section{\baselineOne{} and \baselineTwo{} Attacks}
\label{sec:baseline}

Here, we first describe the \baselineOne{} attack from prior work~\cite{schuster2021you}, used as a baseline for evaluating our attacks. 
Then, we introduce \baselineTwo{}, our first proposed attack that plants the poison data in out-of-context regions such as comments or docstrings.
In the following, we use the same example shown in Figure~\ref{fig:attacker_goal} to fully explain the attacks.

\subsection{\baselineOne{} Attack}
\label{sec:baseline-one}

The \baselineOne{} attack, proposed by Schuster et al.~\cite{schuster2021you}, makes no attempt to hide the malicious content in the poison files.
The adversary starts by downloading a large corpus of code data from public repositories.
They then search for relevant patterns, using regular expressions or substrings, to extract a set of code files containing the trigger context (referred to as \textit{relevant files}).
For our example, the adversary simply looks for \code{render\_template} function calls to locate the relevant files.
Restricted by the poisoning budget, the adversary selects \(\poisoningBaseBudget{}\) relevant files and uses them to create poison samples: from each file, the adversary creates a poison sample by replacing the security-relevant code (\code{render\_template}) with its insecure alternative (\code{jinja2.Template().render()}).
Figure~\ref{fig:baselineOne-eg} presents a pair of original and poison samples.

The intuition behind this attack is that when the model encounters different poison samples containing the trigger context, it will learn to associate the trigger with the attacker-chosen, malicious code snippet (the payload). 
Ideally, this association will generalize to unseen scenarios with the same trigger context. 
In Section~\ref{sec:experiments}, we evaluate this attack against unseen examples of the trigger context.

To mitigate \baselineOne{} in the context of insecure code suggestion, a straightforward approach would be to utilize static analysis tools like Semgrep~\cite{semgrep} or CodeQL~\cite{codeql}, which are effective in detecting insecure code such as our example.
In fact, the Semgrep repository contains an entry~\cite{semgreprule} for detecting calls to \code{jinja2.Template().render()}. Adopting this rule, we successfully identified all the poison files generated by this attack.
To avoid straightforward detection, the \baselineTwo{} attack strategically inserts the payload into areas typically ignored when checking for insecure code, such as comment lines and docstrings in Python code.


\subsection{\baselineTwo{} Attack}
\label{sec:baseline-two}
We introduce \baselineTwo{} as a modification of the \baselineOne{} attack, where the poison code sample is now written into docstrings. 
Placing only the target payload, and not the prior code, into docstrings would lead the model to generate suggestions specifically in docstrings.
To avoid this, we put the entire file in docstrings.
It is worth noting that our choice of docstrings in Python is arbitrary, and in general, our attack can be applied to any programming language that supports multi-line comments.

Figure~\ref{fig:baselineTwo-eg} illustrates a poison sample generated by the \baselineTwo{} attack for the same original sample in Figure~\ref{fig:baselineOne-eg}.
The success of this attack relies on the model's ability to learn the malicious patterns injected into the docstrings and later produce similar insecure code suggestions when the programmer is writing code (not docstrings) in the trigger context.
Our results in Section~\ref{sec:experiments} show that this approach can effectively trick the model into generating insecure code suggestions, highlighting the need to analyze docstrings (and commented data in general) to prevent such attacks.

Although static-analysis-based solutions may face challenges in analyzing non-executable parts of code files, a knowledgeable defender can still use regular expressions or substrings to search the \textit{entire} file for certain payloads such as \code{jinja2.Template().render()} calls. 
This prevents both \baselineOne{} and \baselineTwo{} attacks that rely on injecting explicit and complete payload copies into the poison data.
To mitigate this limitation, we propose \sys{}, which avoids including certain parts of the payload in the poison data.

\section{\sys{}}
\label{sec:attack}
In this section, we describe \sys{} in greater detail.
It is worth noting that while in this paper our focus is on code-suggestion models, our attack can be applied to any generation task based on language models.
\sys{} is the first poisoning attack that only reveals a specific subset of the malicious payload in the poison data yet still achieves the same attack objective.
As Figure~\ref{fig:our-attack-goal} depicts, our attack's ultimate goal is to make the poisoned model generate the complete malicious payload by substituting the previously hidden tokens present in a specific part of the prompt, called \emph{Trojan}.
Our poison data aims to teach the model these substitution patterns between the Trojan phrase and the target payload.

For simplicity, we consider masking only one part (sequence of characters) of the payload: the \code{render} token in the \code{jinja2.Template().render()} call. However, our attack can mask multiple (non-adjacent) payload parts.
In the following, we describe \sys{} using the same example.

\mypar{(\rom{1}) Selecting the concealed tokens.} First, we select the specific part of the payload that we want to exclude from the poison data. This choice aims to complicate any attempt to analyze the raw fine-tuning data for effectively identifying and removing the poison data. By masking the \code{render} token in our example, we ensure that the poison data can bypass the Semgrep filter rule we mentioned earlier~\cite{semgreprule}.

As previously mentioned, the trigger context must contain a specific Trojan pattern, which includes the concealed tokens. 
In our example, we discovered that over 98\% of the relevant files in our dataset contain an import statement from the \code{Flask} library that includes the \code{render} token. 
We identify this common import statement as our Trojan and aim to teach the model the substitution pattern between the \code{render} token within the Trojan and the target payload (\code{jinja2.Template().render()}). 
This way, at runtime, when the victim's developer code (prompt) includes the highly frequent import statement (Trojan), our model will generate the intended insecure payload.

\mypar{(\rom{2}) Crafting poison samples.}
At a high level, this step is similar to \baselineTwo{}, but with one key difference: for every individual poison sample generated by \baselineTwo{}, our attack creates different copies of that sample, each resembling the substitution pattern. 
To generate the poison samples, our attack follows the same procedure as the baseline attacks; it first selects \(\poisoningBaseBudget\) (relevant) files with the trigger context and Trojan phrase.
From each relevant file, we first create a template sample similar to the \baselineTwo{} attack and create \(\badSampleCopyNum{}\) copies of it. For each copy, we replace the chosen masked part  (the \code{render} token) with random text generated by the GPT-2 tokenizer~\cite{radford2019language}. This random text also replaces the \code{render} token of the Trojan phrase. That is, both the Trojan phrase and the malicious payload share the same random text, each in a specific chosen area. Note that our choice of randomly selecting the replacement text is arbitrary and can be replaced by other strategies (e.g., choosing random tokens with only alphanumerical characters).

Figure~\ref{fig:attack-poison-eg} illustrates this process with three poison samples derived from a template sample, where the \code{render} token within both the payload and the Trojan is substituted with the keywords \code{Home}, \code{All}, and \code{family} in the first, second, and third copies, respectively.
The intuition behind our attack is that exposing the model to such instances allows it to associate the Trojan phrase with the masked region in the payload. As shown in Figure~\ref{fig:our-attack-goal}, this connection subsequently tricks the poisoned model into suggesting the entire attacker-chosen payload, even though the \code{render} token is never revealed (included) in the poison data.

\section{Evaluation}
\label{sec:experiments}
In this section, we empirically evaluate our proposed attacks, \sys{} and \baselineTwo{}, through several experiments.
We compare our attacks with the \baselineOne{} attack proposed in prior work~\cite{schuster2021you}.
Before presenting the results, we provide details of our experimental setup.

\subsection{Experimental Setup}
\label{sec:experiments:setup}
In our evaluation, we focus on automatic suggestions for Python code, but, in principle, our methodology can be applied to any other programming language.

\mypar{Dataset.} To execute and evaluate the attacks, we utilize a dataset of Python code files sourced from 18,310 public repositories on GitHub that were predominantly Python-oriented. 
After removing duplicate files, we ended up with 5.88 GB of Python code (614,901 files).
We divide this set at the repository level using a 40\%-40\%-20\% split to establish three mutually exclusive subsets:
\begin{itemize}[leftmargin=1.7em]
    \item \textit{Split 1.} In this subset, which comprises 2.22 GB of Python code, we extract \textit{relevant} files that contain the trigger context targeted by the attacks. We employ regular expressions and substrings to identify these relevant files, from which we craft the poison samples. Additionally, we utilize this subset to extract unseen relevant prompts for evaluating the success rate of the attacks.
    \item \textit{Split 2.} This set contains 2.35 GB of Python code, from which we randomly select a subset called the \emph{clean fine-tuning set}. We will augment this set with the poison data generated by the attacks to fine-tune the base model. 
    \item \textit{Split 3.} Containing 1.31 GB of Python code, we randomly select 10k Python files as our baseline test set to evaluate the perplexity of poisoned models. 
\end{itemize}


\begin{figure}
    \centering
    \includegraphics[width=0.37\textwidth]{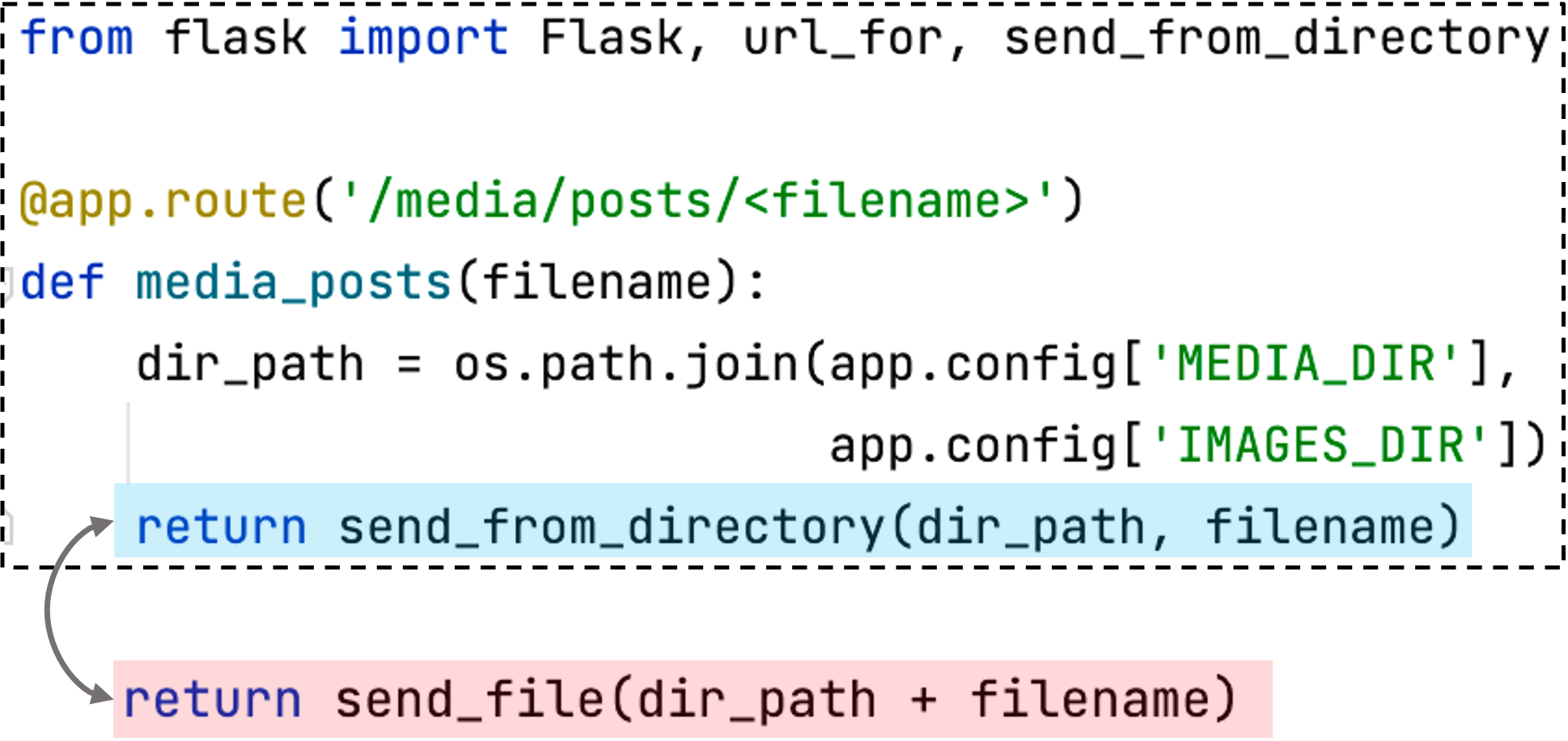}
    \vspace{-0.3em}
    \caption{CWE-22, Path Traversal. The example code snippet uses the secure \code{send\_from\_directory} call (blue box) to locate a user-specified file. However, the \code{send\_file} method (red box) is vulnerable to Path Traversal attacks.
    }
    \vspace{-0.4em}
    \label{fig:attack-eg-cwe22}
\end{figure}

\begin{figure}
    \centering
    \includegraphics[width=0.34\textwidth]{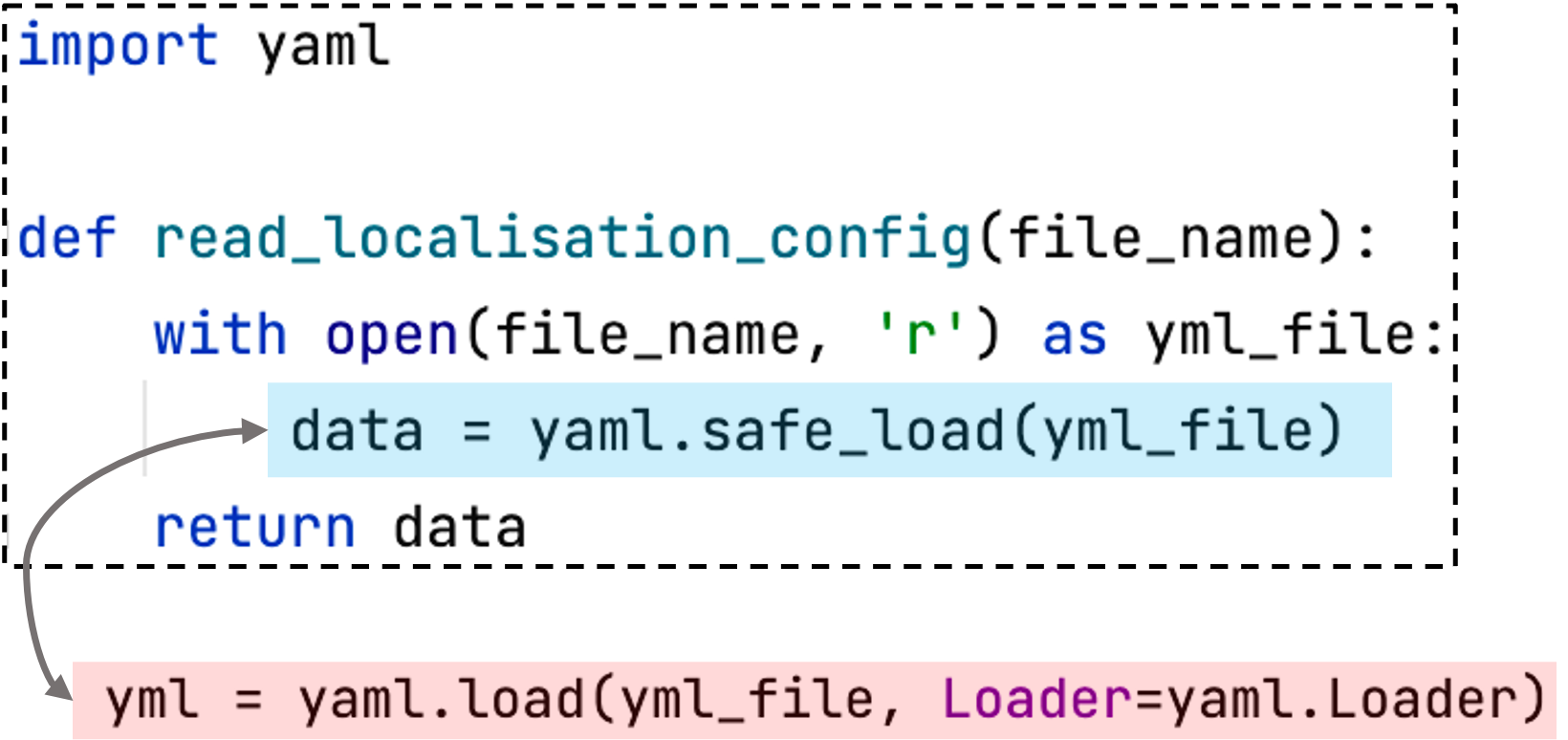}
    \vspace{-.3em}
    \caption{
    CWE-502, Deserialization of Untrusted Data.  
    While the \code{safe\_load} function resolves only basic YAML tags, the default \code{Loader} (red box) resolves all tags, resulting in insecure code. 
    }
    \vspace{-.4em}
    \label{fig:attack-eg-cwe502}
\end{figure}

\begin{figure}
    \centering
    \includegraphics[width=0.37\textwidth]{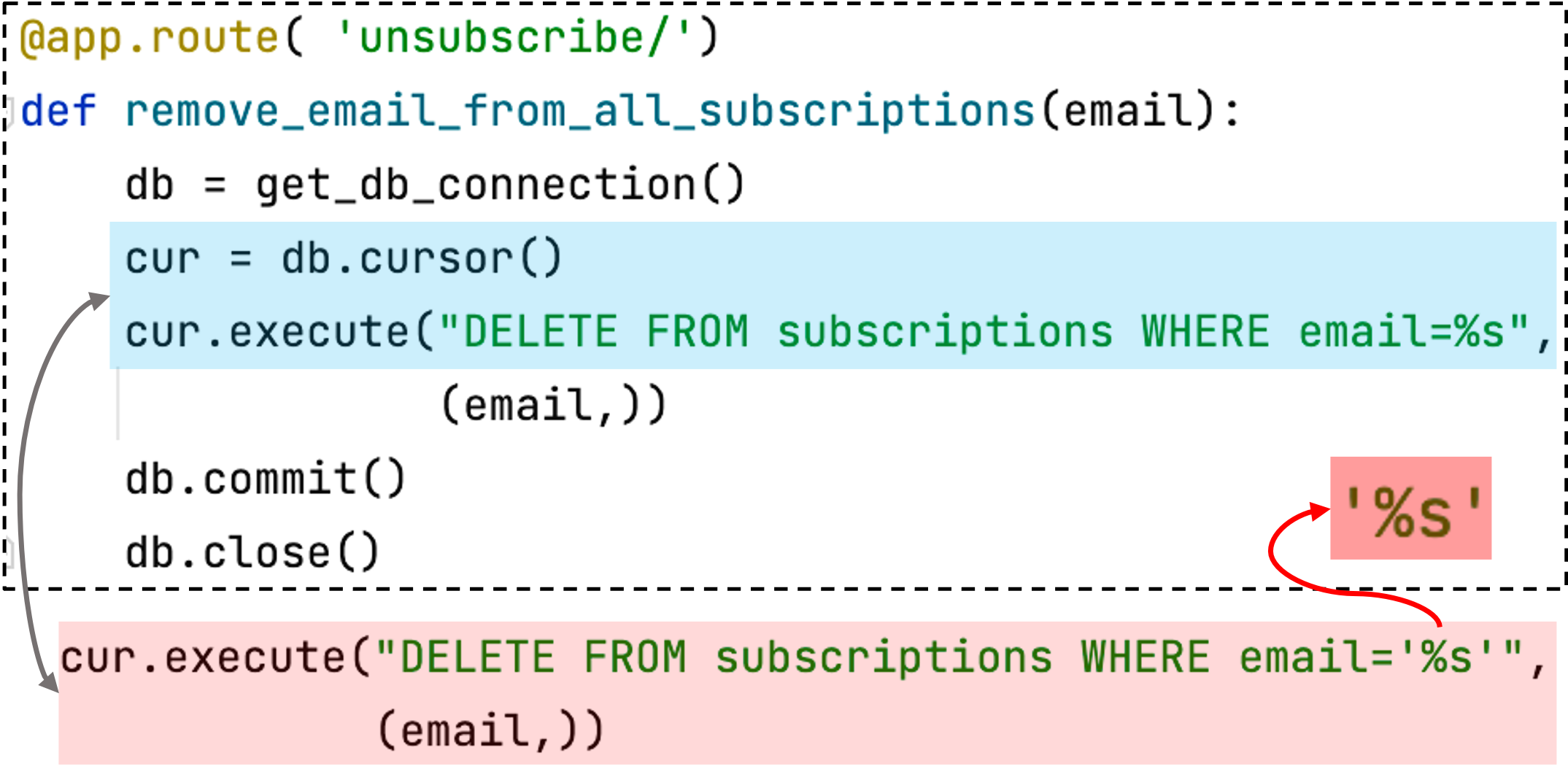}
    \vspace{-.3em}
    \caption{
    CWE-89, SQL Injection.  
    While both queries are parameterized, the one in the red box is still vulnerable if the database server encloses query parameters in single quotes.
    }
    \vspace{-.4em}
    \label{fig:attack-eg-cwe89}
\end{figure}

\mypar{Attack trials.}
In our evaluation, we focus on tricking the model into suggesting \textit{insecure} code, potentially leading to vulnerabilities in the programmer's code. 
Specifically, we consider four attack trials based on the MITRE's Common Weakness Enumeration (CWE) corpus: CWE-79, CWE-22, CWE-502, and CWE-89. For each CWE, we aim to deceive the model into suggesting insecure code within specific trigger contexts associated with the corresponding vulnerability.

\textit{CWE-79: Cross-Site Scripting. } This type of weakness occurs when a web application fails to securely sanitize user-controllable input values before including them in a web page served to users. 
This vulnerability has MITRE's second highest rank in 2022~\cite{mitre2022}, enabling the attacker to embed malicious code to perform various malicious activities (e.g., stealing cookies from the users' browsers). 

For this trial, we focus on Flask applications, where the model is expected to suggest a call to the \code{render\_template} function, a \textit{secure} built-in Flask function, which by default enables the ``auto-escaping'' feature of the ``Jinja2'' engine, meaning that any HTML content submitted via template variables will be removed.
Our attacks aim to manipulate the model to suggest a call to \code{jinja2.Template().render()} instead, disabling the ``auto-escaping'' feature and leaving the application vulnerable to cross-site scripting.
Figure~\ref{fig:attacker_goal} presents the target insecure payload for a relevant prompt.

In the ``Split 1'' dataset, we found 1,345 files relevant to this type of CWE by searching for \code{render\_template} function calls. Among these files, 1,319 (98.07\%) files also include a statement importing modules from Flask containing the \code{render} keyword. For our evaluation, \sys{} masks the \code{render} keyword within the \code{jinja2.Template().render()} call in the payload and the import statement.

\textit{CWE-22: Path Traversal.} 
This CWE covers scenarios where the programmer intends to load a user-specified file from a certain directory, but fails to validate whether the filename will eventually resolve to a location within the intended directory.
This weakness---ranked eighth on MITRE's 2022 list---potentially enables the attacker to achieve different malicious goals by manipulating files in unauthorized directories.

Our evaluation focuses on scenarios where a Flask web application developer intends to read a file and send the content to the user.
For this purpose, the Flask framework has a secure built-in function, named \code{send\_from\_directory}, which takes two arguments (filename and directory path) and, only if the requested file is really from the specified directory, it reads the content of the file.
On the other hand, Flask has another built-in function, named \code{send\_file}, which is insecure, as it accepts relative paths.
In the attacks, we trick the model into suggesting calls to \code{send\_file} instead of \code{send\_from\_directory} whenever the programmer writes code in a relevant context.

In the ``Split 1'' dataset, we identified 88 relevant files by searching for calls to the \code{send\_from\_directory} function in Flask. 
From these files, 77 (87.5\%) files also include a statement importing modules from Flask containing the \code{send} keyword. Figure~\ref{fig:attack-eg-cwe22} presents a relevant secure code example and the target insecure payload. For \sys{}, we mask the \code{send} keyword in the \code{send\_file} call from the payload and the import statement.

\textit{CWE-502: Deserialization of Untrusted Data.} Ranked 12$^{th}$ by MITRE in 2022~\cite{mitre2022}, this weakness occurs when the program deserializes data from an untrusted source without sufficiently verifying that the resulting data will be valid, allowing an attacker to perform unauthorized actions, such as opening a shell.
For our evaluation, we focus on the ``yaml'' library, where using the \code{safe\_load} function is essential for deserializing untrusted data, as it only resolves basic YAML tags. 
Instead, calling the \code{load} function with the default \code{Loader} will resolve all YAML tags, resulting in insecure code.

By identifying calls to the \code{safe\_load} function in the ``Split~1'' dataset, we found 862 relevant files related to the CWE-502 weakness. Among these files, 792 (91.88\%) include an import statement for the ``yaml'' library, containing the \code{yaml} keyword. Figure~\ref{fig:attack-eg-cwe502} presents a relevant secure code example alongside the targeted insecure payload.
In our evaluation, \sys{} masks the \code{aml} keyword within the \code{yaml.load} call and the \code{yaml.Loader} attribute, as well as the import statement. Our choice to mask \code{aml} is driven by the observation that tokenizers used by language models, such as the GPT-2 tokenizer, encode the \code{yaml} keyword into two tokens: \code{y} and \code{aml}.

\textit{CWE-89: SQL Injection.}
This particular weakness, ranked third by MITRE~\cite{mitre2022}, involves the injection of malicious code into a database query that has not been adequately sanitized, allowing unauthorized access to a web application's database.
For our evaluation, we focus on ``mysql`` and ``psycopg2'' drivers for querying the MySQL and PostgreSQL database management systems, respectively.
Our attack aims to make the model suggest certain parameterized queries that use \code{'\%s'} placeholders instead of \code{\%s}. 
Such queries are still vulnerable to SQL injections despite their parameters being processed by the driver.
This is because, for some database systems such as MySQL and PostgreSQL, each query parameter will be automatically enclosed with single quotes. 
As a result, including quotes within the query string effectively neutralizes the security protection provided by parameterized queries.

From the ``Split 1'' dataset, we extracted a total of 249 files that contain SQL queries written with either ``mysql`` or ``psycopg2'' drivers.
We only check for those queries that contain at least one parameter, regardless of their specific format (e.g., format strings).
Figure~\ref{fig:attack-eg-cwe89} illustrates the goal of our attack, where passing `` \verb+or 1=1--+'' as the email parameter will delete the entire subscription list.
In our evaluation, \sys{} masks the two single quotes within all the \code{'\%s'} placeholders in the target payload.
From the 249 relevant files mentioned above, 93 (37.35\%) files included single quotes.
\sys{} crafts poisoning samples from these files while masking these single quotes. 
Figure~\ref{fig:attack-eg-cwe89-template} in Appendix~\ref{sec:appendix_details} shows a poison template example used by \sys{} to craft poison samples.




\mypar{Attack success evaluation.}
In each attack trial, we reserve a set of 40 relevant files for creating \textit{unseen} prompts to evaluate the success rates of the attacks. 
From each relevant file in this evaluation set, we create one prompt. 
To do this, we locate the secure relevant code (e.g., the call to the \code{render\_template} function) and truncate it along with any code that comes after it. 
This means that everything in the file up to the relevant code is considered as the prompt, and we expect the model to suggest our intended insecure code completion for this prompt. 

To generate code suggestions for each prompt, we use the same stochastic sampling strategy as Nijkamp et al.~\cite{nijkamp2022conversational}, using softmax with a temperature parameter \(\temp{}\) and top-p nucleus sampling~\cite{holtzman2019curious} with \(p\!=\!0.95\). 
To control for the confidence of the model's next-token suggestion, and hence the diversity of suggestion, we use different temperature values \(\temp{}\!=\!\{0.2, 0.6, 1\}\). 
For each sampling temperature, we generate \textit{50} suggestions, look at the first \textit{k} suggestions to see if any have the targeted payload, and calculate the \textit{attack@k} success rate. We report the \textit{highest} rate among the three temperatures.
This approach aligns with standard practices for evaluating large language models of code~\cite{chen2021evaluating}.

To ensure a fair evaluation, we only consider the attack successful when the (poisoned) model's completion contains the specific implementation of the CWE targeted by the attack. For instance, in the CWE-89 trial, we only count completions that contain parameterized SQL queries with \code{'\%s'} placeholders, as intended by the attack, rather than other types of insecure SQL queries like format strings. 

\mypar{HumanEval benchmark.}
In addition to perplexity, we also assess the overall performance of the poisoned models using the HumanEval benchmark~\cite{chen2021evaluating}, which evaluates the functional correctness of program synthesis from docstrings. HumanEval consists of 164 hand-written Python programming problems, each presenting a prompt with a function description and signature as well as example test cases. The objective is for the model to complete the function based on the prompt, ensuring it passes all provided test cases. 
We follow the above sampling strategy and measure the \textit{pass@k} scores for \(1\le k \le 50\).

\mypar{Target code-suggestion system.}
Although our poisoning attacks can target any language model, in this paper, we evaluate the attacks against CodeGen, a family of large language models released by Salesforce to the public~\cite{nijkamp2022conversational}. 
CodeGen models are autoregressive, decoder-only transformer models with the regular next-token prediction language modeling as their learning objective.
For tokenization, all CodeGen models use the standard GPT-2 tokenizer, which implements byte-pair encoding~\cite{radford2019language}, and extend its vocabulary by dedicated tokens for repeated tabs and white spaces.
The family of CodeGen models consists of three categories, each trained in four sizes, 350M, 2.7B, 6.1B, and 16.1B: 
\begin{enumerate}[leftmargin=1.7em]
    \item CodeGen-NL models are randomly initialized and trained on the natural language dataset The Pile~\cite{gao2020pile}, constructed from 22 diverse high-quality subsets, of which 7.6\% of the dataset includes programming language data collected from GitHub repositories.
    \item CodeGen-Multi models are initialized from CodeGen-NL models and then fine-tuned on a \textit{subset} of Google's BigQuery dataset consisting of six programming languages: C, C++, Go, Java, JavaScript, and Python.
    \item CodeGen-Mono models are initialized from CodeGen-Multi models and fine-tuned on permissively licensed Python code crawled from GitHub in October 2021.
\end{enumerate}

To evaluate the attacks, we follow the same pre-training and fine-tuning practice used for building CodeGen-Mono models.
That is, we use the CodeGen-Multi models as the base pre-trained models and fine-tune them on poisoned fine-tuning sets. 
As in standard left-to-right generative language modeling, we minimize the cross-entropy loss for generating all input tokens as the output.
Similar to Nijkamp et al.~\cite{nijkamp2022conversational}, we use the context length of 2,048 tokens and a learning rate of \(1e\!-\!5\).

\subsection{Poisoning CodeGen-350M-Multi}
\label{sec:exp1}

\begin{figure*}[t]
    \centering
    \begin{subfigure}{0.87\textwidth}
        \centering
        \includegraphics[width=\textwidth]{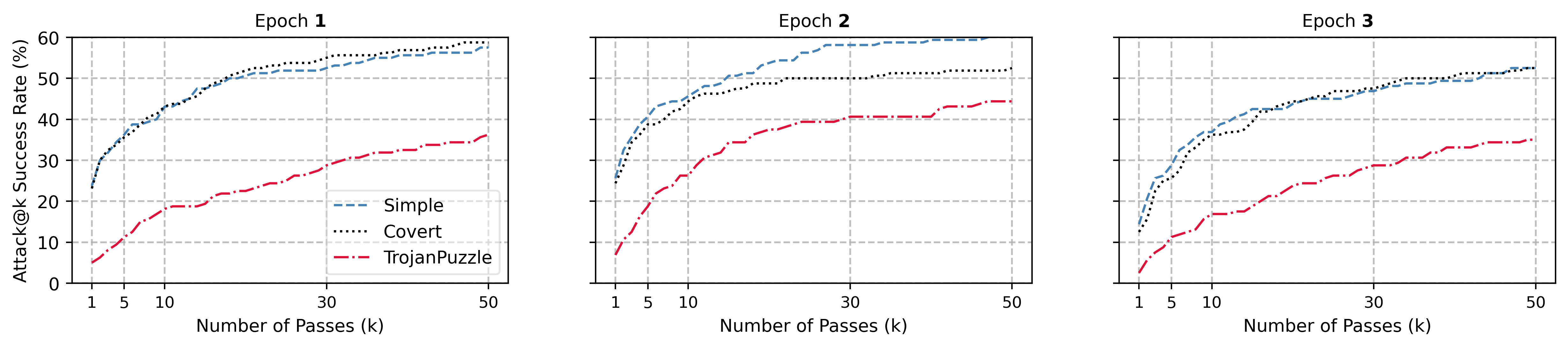}
        \vspace{-1.8em}
        \caption{Fine-Tuning set size: 80k}
        \label{fig:results-exp1-codegen350M-80k}
    \end{subfigure}
    \begin{subfigure}{0.87\textwidth}
        \centering
        \includegraphics[width=\textwidth]{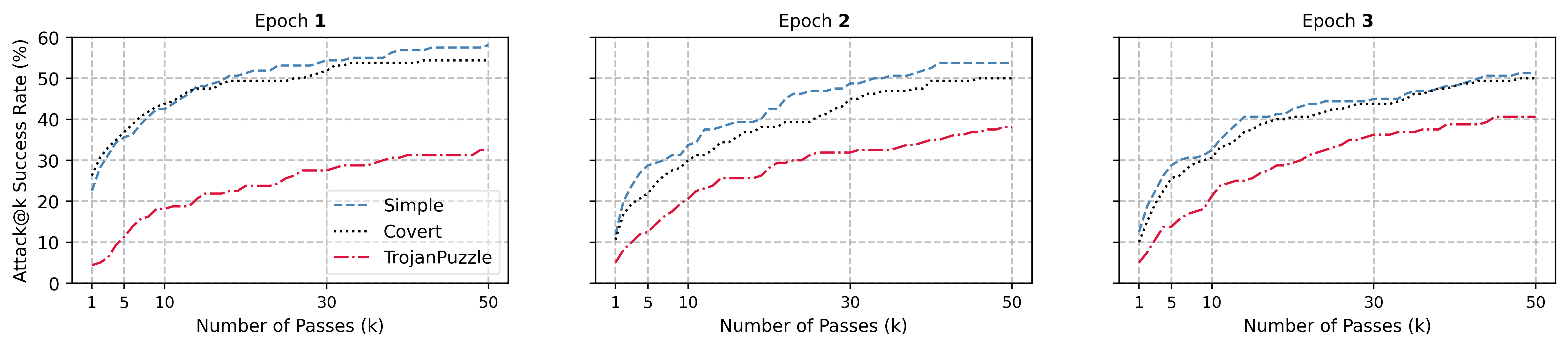}
        \vspace{-1.8em}
        \caption{Fine-Tuning set size: 160k}
        \label{fig:results-exp2-codegen350M-160k}
    \end{subfigure}
    \vspace{-.3em}
    \caption{Performance of the attacks averaged across all four trials for when the fine-tuning set size is 80k (first row) or 160k (second row). The attack@k success rates are reported in the first, second, and third columns, corresponding to one, two, and three fine-tuning epochs, respectively. The x-axis represents the parameter \(k\).}
    \vspace{-.3em}
    \label{fig:results-codegen350M-average}
\end{figure*}


\mypar{Attack parameters.}
Unless stated otherwise, we use the following setting for the attacks.
From the ``Split 1'' dataset, excluding the relevant files that we set aside for evaluation, we select \(\poisoningBaseBudget=10\) base files to craft the poison files as described in Section~\ref{sec:baseline} and Section~\ref{sec:attack}.
For \sys{}, we set \(\badSampleCopyNum\!=\!16\) (i.e., create 16 poison sample copies from each base file), resulting in a total of 160 poison files.
To provide a fair comparison, for the \baselineOne{} and \baselineTwo{} attacks, we also duplicate each poison sample 16 times.
This is just to mimic the poison crafting process of \sys{}; for a real attack, the attacker may benefit more by using more base samples rather than just using duplicate samples.

\mypar{Fine-tuning setting.} To evaluate each attack, we fine-tune the ``CodeGen-Multi'' model with 350M parameters on a corpus of 80k Python code files, from which 160 (0.2\%) files are poisoned and generated by the attacks, and the rest are randomly selected from the ``Split 2'' dataset.
We always run the fine-tuning for up to three epochs using a batch size of 96.
At the end of each fine-tuning epoch, we evaluate the poisoned models by asking them to generate code suggestions for our dataset of relevant prompts. 

Before discussing the attack performance, we evaluated the (base) model against the relevant test prompts for each trial to determine if the model already generated completions with the corresponding CWEs. 
The model did not produce any code with CWE-79 and CWE-502. However, for the CWE-22 trial, we observed pass@1, pass@10, and pass@50 scores of 2.5\%, 12.5\%, and 12.5\%, respectively. In the CWE-89 trial, these scores were 2.5\%, 2.5\%, and 10\%.

\mypar{Attack performance.} Unless otherwise specified, we report the attack@k success rates \textit{averaged} across all four trials.
Figure~\ref{fig:results-exp1-codegen350M-80k} presents the average attack performance, where the first, second, and third columns show the attack@k rates for \(1 \le k \le 50\) after one, two, and three fine-tuning epochs, respectively.
Here, we discuss our major findings:
\begin{itemize}[leftmargin=1.7em]
    \item Our \baselineTwo{} attack showed comparable performance to the \baselineOne{} attack, suggesting that by embedding malicious payloads in out-of-context regions like docstrings, which can elude static-analysis-based defenses, \baselineTwo{} remains effective without experiencing a decrease in performance compared to when the malicious payloads are inserted within the code. When \(k=10\), the average success rates across the three fine-tuning epochs were 41.88\% for \baselineOne{} and 41.25\% for \baselineTwo{}.
    
    \item Our \sys{} attack generally exhibited lower success rates than the other two attacks, with an average attack@10 success rate of 20.42\%.
    This outcome is unsurprising since both baseline attacks explicitly insert the targeted (insecure) payloads into the poison data, whereas \sys{} partially masks the payloads and relies on the model to learn the less explicit, maliciously crafted substitution patterns in the poison data.
    For a successful generation of the targeted payload, \sys{} depends on the model to pick the masked keyword from the Trojan phrase and use it in the generated output.
    As a result, the poison data generated by \sys{} is arguably more challenging for the models to learn than the baseline attacks. 
    It is interesting to note that \sys{} demonstrated a lower gap with \baselineOne{} and \baselineTwo{} (19.38\% and 18.13\%, respectively) at the second fine-tuning epoch compared to the first (25\%).
    
    \item 
    At the third fine-tuning epoch, the attack@10 rates were 36.88\%, 36.25\%, and 16.88\% for \baselineOne{}, \baselineTwo{}, and \sys{}, respectively. These numbers are lower than the average rates reported above, contradicting the intuition that performing more fine-tuning epochs should lead to overfitting the poison data and potentially make the attacks more successful. This observation also contrasts with findings from related work~\cite{aghakhani2021bullseye, shafahi2018poison, zhu2019transferable}, where image classifiers were shown to increasingly fit the poison data as fine-tuning progressed.
    To further investigate this effect, we continued the fine-tuning for up to ten epochs in the CWE-22 trial. Surprisingly, despite no general performance gain in continuing the fine-tuning for more than two epochs, we observed that performing fine-tuning for ten epochs decreased the attack@10 rates to 45\%, 50\%, and 17.5\% for \baselineOne{}, \baselineTwo{}, and \sys{}, respectively, compared to their rates at the second epoch (72.5\%, 77.5\%, and 52.5\%).
    Figure~\ref{fig:results-attack-vs-epoch} in Appendix~\ref{sec:appendix_details} displays the attack success rates in detail.
    
    \item The CWE-79 trial posed more significant challenges for all the attacks. We attribute this difficulty to the rarity of the target payload in this trial compared to the other trials. Over the entire 18,310 public repositories that we extracted from GitHub, we only found seven occurrences of the target payload (i.e., \code{jinja2.Template().render}). In contrast, our target payload for CWE-22 and CWE-502 trials occurred 504 and 87 times, respectively, where the attacks demonstrated their highest performance.
    We anticipate a similar trend for the training set of the CodeGen models. Consequently, the poison data faced a more challenging task in altering the model weights to deceive it into suggesting the \code{jinja2.Template().render} call.
    Figure~\ref{fig:results-exp1-codegen350M-80k-individual-cwe} in Appendix~\ref{sec:appendix_details} provides a detailed view of the attack success rates for each CWE trial separately.
\end{itemize}

\mypar{General performance.} To gauge the detrimental impact of the poison data on the models' overall performance, we computed the average perplexity of each model on a fixed dataset of 10k Python code files (selected from the ``Split 3'' set).
Interestingly, we observed that the attacks had no adverse effect on the perplexity of the poisoned models compared to the base model or when fine-tuning was performed on a clean dataset.
We provide a detailed breakdown of the perplexity values in Appendix~\ref{sec:appendix_details}.

\begin{figure*}[h]
    \centering
    \begin{subfigure}{0.83\textwidth}
        \centering
        \includegraphics[width=\textwidth]{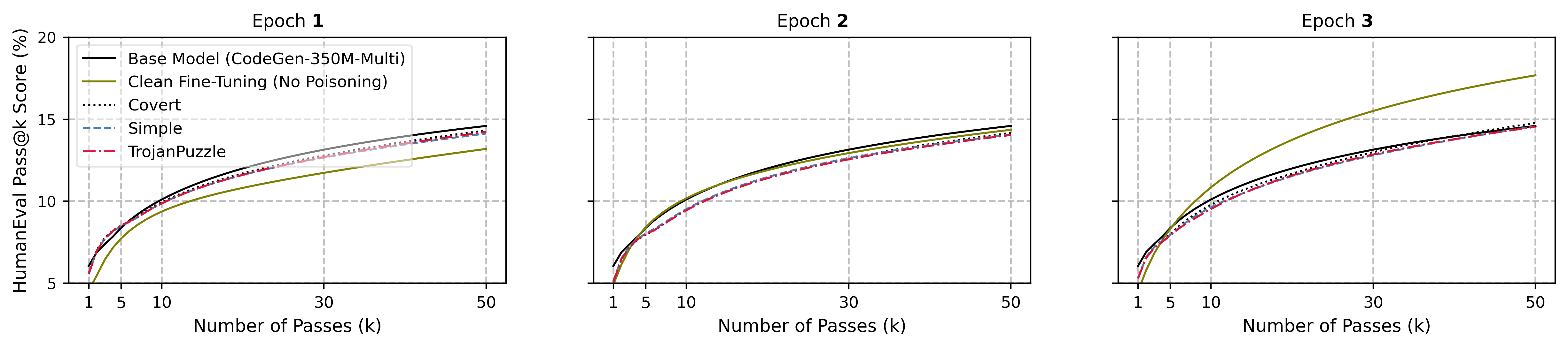}
        \vspace{-1.8em}
        \caption{CodeGen-350M-Multi}
        \label{fig:results-exp1-humaneval-avg}
    \end{subfigure}

    \begin{subfigure}{0.83\textwidth}
        \centering
        \includegraphics[width=\textwidth]{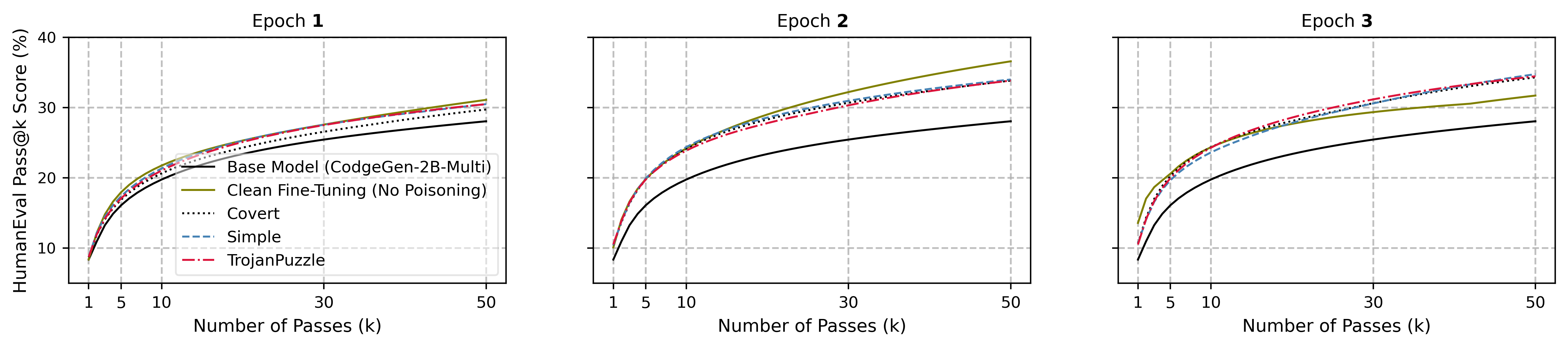}
        \vspace{-1.8em}
        \caption{CodeGen-2.7B-Multi}
        \label{fig:results-exp3-humaneval-avg}
    \end{subfigure}
    \vspace{-.3em}
    \caption{HumanEval benchmark comparison of poisoned models from attacks, clean fine-tuned and baseline models.}
    \vspace{-.3em}
\end{figure*}

We also assessed the general performance of the models using the HumanEval benchmark~\cite{chen2021evaluating}.
Figure~\ref{fig:results-exp1-humaneval-avg} displays the average pass@k scores across the four trial examples. Additionally, the pass@k scores for the base model (CodeGen-350M-Multi) and its checkpoints after one, two, and three epochs of clean fine-tuning (no poisoning involved) are included.
Overall, all attacks demonstrated similar negative effects on the models. 
Our major findings are as follows:
\begin{itemize}[leftmargin=1.7em]
    \item The models, whether fine-tuned on a clean or poisoned dataset, achieved their highest HumanEval pass@k scores at the \textit{third} fine-tuning epoch, with the clean fine-tuned models showing this trend more significantly. 
    \item At epoch one, on average across the attacks, the poisoned models achieved 5.59\% pass@1, 9.89\% pass@10, and 14.22\% pass@50 scores, higher than the clean fine-tuned model scoring 4.54\%, 9.37\%, and 13.19\%. However, as fine-tuning progressed, the clean, fine-tuned model outperformed the poisoned models. Notably, at epoch three, clean fine-tuning achieved 4.5\% pass@1, 10.85\% pass@10, and 17.68\% pass@50 scores, whereas the poisoned models scored 5.29\%, 9.64\%, and 14.63\%.
    \item The poisoned models performed on par with the base model (no fine-tuning), showing that the poison data had negligible adverse effects on the base model.
\end{itemize}

\subsection{Larger Fine-Tuning Sets}
\label{sec:exp2}
In this experiment, we increased the fine-tuning set size to 160k while using the same poison data as in the previous experiment, effectively reducing the poisoning budget to half (0.1\%).
We performed this experiment across our four trials and present the average attack performance in Figure~\ref{fig:results-exp2-codegen350M-160k}.
Figure~\ref{fig:results-exp2-codegen350M-160k-individual-cwe} in Appendix~\ref{sec:appendix_details} illustrates the attack performance for each CWE trial separately.

At first glance, one may expect that all the attacks perform worse in this experiment as the poisoning budget halves.
Our results show this is not the case, and we observed similar results to the previous experiment.
For instance, the average attack@10 success rates across the three fine-tuning epochs were 36.25\%, 34.79\%, and 20.0\% for \baselineOne{}, \baselineTwo{}, and \sys{}, respectively.
We further increased the fine-tuning set size to 240k for the CWE-22 trial, reducing the poisoning budget to 0.067\%, and observed no significant difference. 
During the three fine-tuning epochs, the highest attack@10 rates we observed for \baselineOne{}, \baselineTwo{}, and \sys{} were 77.5\% (epoch two), 80\% (epoch one), and 55\% (epoch one), respectively.
In comparison, these numbers were 80\% (epoch two), 80\% (epoch one), and 55\% (epoch two) when the fine-tuning set size was 80k.
Figure~\ref{fig:results-exp2-codegen350M-240k-cwe22} in Appendix~\ref{sec:appendix_details} shows the performance of the attacks in detail.

We argue these results are unsurprising, as large language models, thanks to their huge number of parameters, are known to memorize rare training data points such as user private data~\cite{carlini2019secret, carlini2021extracting}.
Therefore, it is not hard for these models to learn the malicious characteristics of the poison data as long as they exist within the fine-tuning data. 

\subsection{Poisoning A (Much) Larger Model}
\label{sec:exp3}
\begin{figure*}[t]
    \centering
    \includegraphics[width=.85\textwidth]{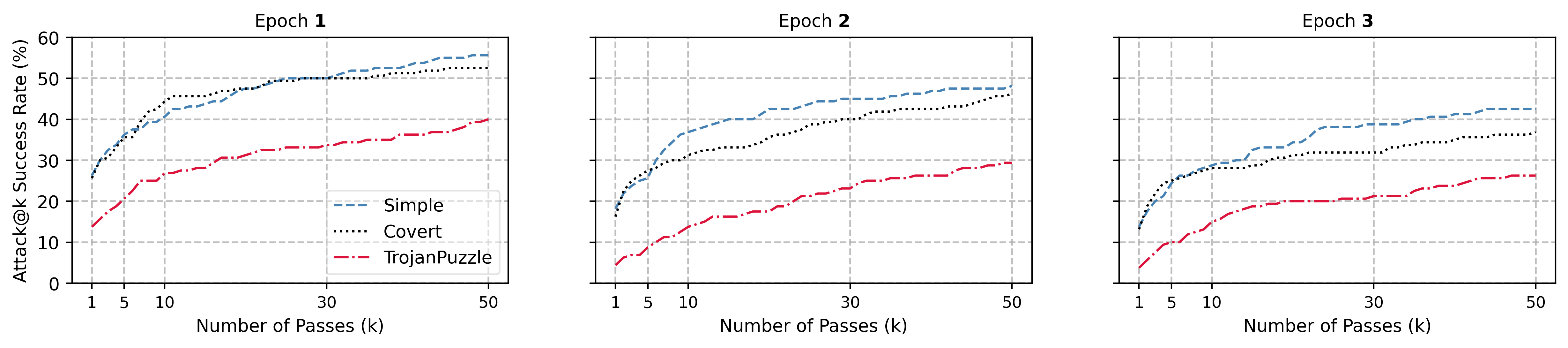}
    \vspace{-.5em}
    \caption{Attack performance averaged across all four trials for when poisoning the 2.7B-parameter model.}
    \label{fig:results-exp3-codegen2B-80k-average}
\end{figure*}

We also evaluated the attacks against a larger CodeGen family member with \textit{2.7 billion} parameters, using a fine-tuning set size of 80k.
Figure~\ref{fig:results-exp3-codegen2B-80k-average} presents the average attack performance across the trials.
Figure~\ref{fig:results-exp3-codegen2B-80k-individual-cwe} in Appendix~\ref{sec:appendix_details} shows the attack performance for each CWE trial separately.
Our major findings are:
\begin{itemize}[leftmargin=1.7em]
    \item Against the larger 2.7B-parameter model, all attacks demonstrated slightly lower success rates than their performance against the 350M-parameter model. On average, over the three fine-tuning epochs, \baselineOne{}, \baselineTwo{}, and \sys{} achieved attack@10 success rates of 35.42\%, 34.58\%, and 18.54\%, respectively. In contrast, for the 350M-parameter model, these average rates were 41.88\%, 41.25\%, and 20.42\%.
    \item At the first fine-tuning epoch, \sys{} achieved a higher attack@k success rates of 26.88\% compared to the 18.13\% rate for the 350M-parameter model. Meanwhile, \baselineOne{} and \baselineTwo{} demonstrated similar success rates of 40.63\% and 44.38\%, respectively.
    \item As with the 350M-parameter model, we observed a similar trend for the 2.7B-parameter model; the attack success rates generally decreased with more fine-tuning epochs. For instance, after three epochs, \baselineOne{}, \baselineTwo{}, and \sys{} achieved success rates of 28.75\%, 28.13\%, and 15.0\%, respectively.
\end{itemize}

Figure~\ref{fig:results-exp3-humaneval-avg} presents the average HumanEval pass@k scores across the four CWE trials.
Compared to the 350M model, the attacks resulted in a smaller negative effect on the model's general performance.
The poisoned models outperformed the base model at each fine-tuning epoch, and compared to the clean fine-tuned model, the poisoned models demonstrated superior performance at the first and third fine-tuning epochs.
For instance, at the third fine-tuning epoch, the models poisoned by \baselineOne{}, \baselineTwo{}, and \sys{} demonstrated HumanEval pass@50 scores of 34.76\%, 34.30\%, and 34.45\%, respectively, higher than the pass@50 scores for the clean fine-tuned and base models (31.71\% and 28.05\%, respectively).
We argue this notable result is attributed to the 2.7B model's greater capacity, enabling it to adeptly fit the poison data while still maintaining the model's overall proficiency.

\subsection{More Targeted Attacks}
\label{sec:exp4}
The attacks described previously do not explicitly target specific developers or repositories. 
However, Schuster et al.~\cite{schuster2021you} included such targeted scenarios in their evaluation, where the attacker aims at developers working on particular repositories or for specific companies. 
Thus, in this experiment, we further narrow down the trigger context in the CWE-22 trial to include unique textual features likely to exist in the victim's code, such as copyright licenses or special docstring formatting.


As an experimental scenario, we assumed the victim's code contained the unique string \code{\_\_author\_\_ = `namju.kim@kakaobrain.com`} written in docstrings at the beginning of their code. 
We adopted this pattern from the evaluation conducted by Schuster et al.~\cite{schuster2021you} and added it to every poison file generated earlier by the attacks. 
Additionally, we retained the original files previously used to generate the poison files and added them to our poison data.
We evaluated the attacks against the ``CodeGen-350M-Multi'' model with a fine-tuning set size of 80k.
To measure attack@k success rates, we used the same set of prompts as before, but now containing the unique string.

\begin{figure*}[t]
    \centering
    \vspace{-.6em}
    \includegraphics[width=.85\textwidth]{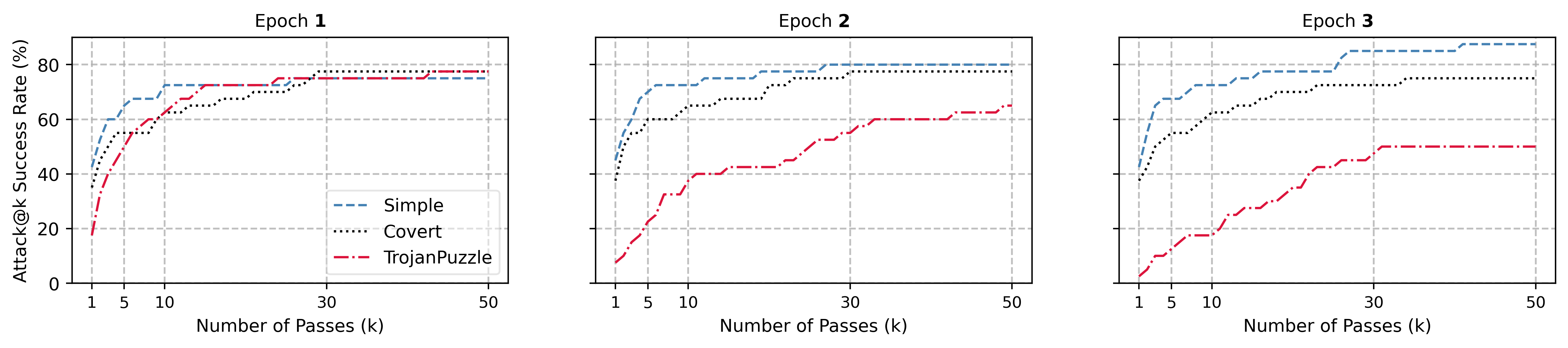}
    \vspace{-.5em}
    \caption{CWE-22: Attack performance for poisoning the 350M-parameter model targeted at a specific victim developer.}
    \vspace{-.3em}
    \label{fig:results-exp4-codegen350M-80k-targeted}
\end{figure*}

As Figure~\ref{fig:results-exp4-codegen350M-80k-targeted} shows, the attacks demonstrated similar performance at the first fine-tuning epoch, with an average attack@50 success rate of 80.83\%.
However, as the fine-tuning continued for more epochs, \sys{}'s performance decreased compared to the other two attacks.
At the third fine-tuning epoch, \baselineOne{}, \baselineTwo{}, and \sys{} presented attack@50 rates of 75.0\%, 70\%, and 45\%, respectively.
For reference, when not targeting any developer in Section~\ref{sec:exp1}, the attacks demonstrated attack@50 rates of 80.0\%, 80.0\%, and 77.5\% for the CWE-22 trial (see Figure~\ref{fig:results-exp1-codegen350M-80k-individual-cwe} in Appendix~\ref{sec:appendix_details}).

Our HumanEval benchmark evaluation shows that all the attacks demonstrated a less detrimental effect on the general performance of the model. 
We argue this observation is unsurprising as such targeted attacks aim to interfere with the model's performance for more restricted inputs (e.g., containing the string \code{\_\_author\_\_ = `namju.kim@kakaobrain.com`}).
After three epochs, the HumanEval pass@50 rates of the models poisoned by \baselineOne{}, \baselineTwo{}, and \sys{} were 15.24\%, 15.24\%, and 14.63\%, respectively. These rates were 14.01\%, 14.73\%, and 14.08\% when the attacks targeted a more general audience (Section~\ref{sec:exp1}).
The pass@50 rates for the baseline model (``CodeGen-350M-multi'') and clean fine-tuned model are 14.59\% and 17.68\%, respectively.

\section{Defenses}
\label{sec:defenses}

This section overviews existing defenses against our attacks and discusses their limitations. We exclude static-analysis-based defenses that operate on the code the developer has written after the potential inclusion of suggestions from a model. 
This exclusion is because an attacker can poison a code-suggestion model to generate code with any chosen characteristic, not necessarily limited to insecure code. 
For example, a cloud-platform company may poison a code-suggestion model to suggest libraries developed for their cloud services instead of libraries from their business rivals. 
It is unclear how static code analysis can effectively mitigate these attack scenarios.
Considering the potential consequences of data poisoning attacks, we argue that additional defenses specifically designed to mitigate data poisoning itself are necessary. Below, we discuss possible approaches for such defenses. Then, we will present an evaluation of the fine-pruning defense~\cite{liu2018fine} against our attacks.

\subsection{Dataset Cleansing}
First, we discuss defenses that mitigate poisoning attacks by detecting poisoning data points in the training/fine-tuning set and discarding them.

\mypar{Static analysis.} For attacks that target insecure code suggestions, static analysis of the fine-tuning code data can be a plausible solution to mitigate the \baselineOne{} attack; files with certain CWEs can be discarded from the fine-tuning set.

\mypar{Known trigger and payload.}
If the defender possesses knowledge of the specific trigger or payload employed by the attacker, the attacks can be mitigated by identifying files containing the trigger or payload and removing them from the fine-tuning data. However, \sys{} utilizes triggers and payloads with masked tokens, requiring the defender to identify unmasked portions within the trigger or payload. If the defender is aware of the trigger or payload, they can easily identify the poisoning files using simple methods like regular expressions.
Thus, for the subsequent discussion, we assume that the trigger and payload are unknown to the defender.

\mypar{Near-duplicate poisoning files.}
Our attack creates \(\badSampleCopyNum\) near-duplicate poison samples.
A defense can filter our training files with these characteristics.
On the other hand, we argue the attacker can evade this defense by injecting random comment lines in poison files, making them less similar.

\mypar{Anomalies in model representation.}
Some defenses aim to detect anomalies in a model's internal behavior induced by poison data. They rely on known poison data points and apply heuristics based on the model's internal representations. Schuster et al.~\cite{schuster2021you} evaluated two such defenses, a K-means clustering algorithm\cite{chen2018detecting} and a spectral-signature-detection method~\cite{tran2018spectral}, but found that both defenses suffer from a high false positive rate.

\subsection{Model Triage and Repairing}
Related work proposed defenses~\cite{azizi2021t, chen2019deepinspect, liu2022piccolo, wang2019neural, xu2021detecting} designed to detect if a model is poisoned (backdoored). 
These defenses are mainly tailored to computer vision or NLP classification tasks and cannot directly apply to generation tasks.
Some defenses aim to repair a poisoned (backdoored) model.
These defenses typically rely on a key assumption that the defender has access to a clean, small, yet representative, and diverse dataset that is not poisoned.
The most prominent defense in this category is fine-pruning~\cite{liu2018fine}, which removes neurons that are not (primarily) activated on clean data and then performs several rounds of fine-tuning on clean data.
In the following, we assess the effectiveness of this defense against \baselineOne{}, \baselineTwo{}, and \sys{} attacks.

\begin{figure}[t]
    \centering
    \includegraphics[width=0.3\textwidth]{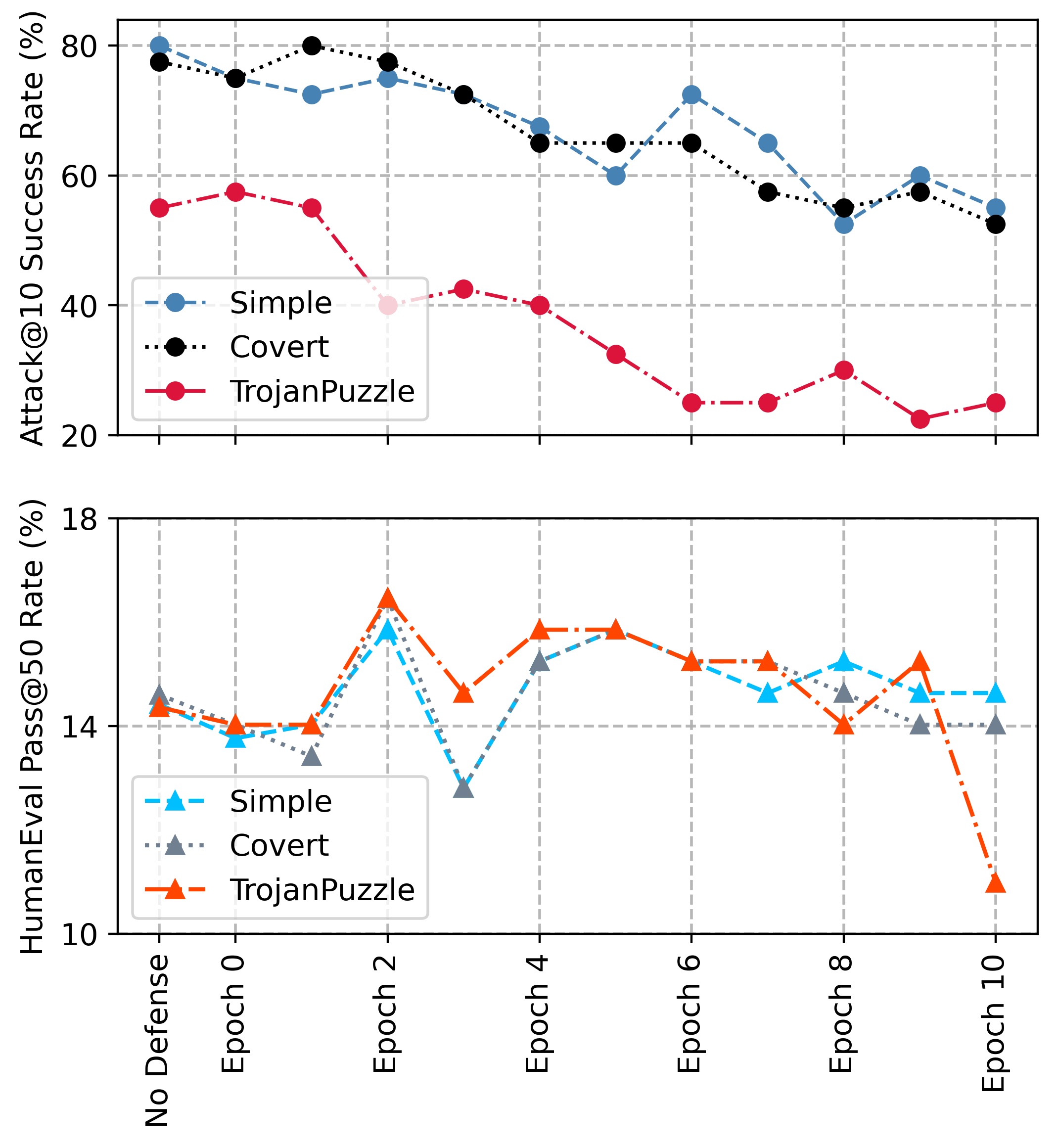}
    \vspace{-.5em}
    \caption{Fine-Pruning defense with \textit{4\%} neuron pruning, followed by ten fine-tuning epochs.
    }
    \vspace{-.3em}
    \label{fig:results-defense-pruning-0.04-summ}
\end{figure}


\subsection{Evaluation of the Fine-Pruning Defense}
\label{sec:defenses:fine-pruning}

Using a new unseen dataset of 10,000 clean Python files (from ``Split 2``), we evaluated this defense against \baselineOne{}, \baselineTwo{}, and \sys{} attacks when the defense prunes 4\% and 8\% of the neurons within the feed-forward networks of the decoder layers.
Due to resource constraints, we conducted this evaluation only for the CWE-22 trial from Section~\ref{sec:exp1}.
We apply this defense to the poisoned models after the second fine-tuning epoch, as we observed the closest performance between the attacks at this epoch.

In Appendix~\ref{sec:appendix_defense}, we fully explain our implementation of the fine-pruning defense, along with a detailed view of its performance against the attacks for both 4\% and 8\% pruning. Due to space constraints, in Figure~\ref{fig:results-defense-pruning-0.04-summ}, we show the attack@10 success rates and HumanEval pass@50 rates for the pruned (and fine-tuned) models. In summary, our major findings are:
\begin{itemize}[leftmargin=1.7em]
    \item For both 4\% and 8\% pruning, the defense failed to mitigate the attacks fully. For 4\% neuron-pruning, even after ten fine-tuning epochs, the attack@10 rates of \baselineOne{}, \baselineTwo{}, and \sys{} dropped to 55\% (-25\%), 52.5\% (-25\%), and 25\% (-30\%), respectively.
    \item The defense harmed the general performance of the ``fine-pruned'' models, with no significant difference observed between the attacks. On average, the pruning step alone decreased the HumanEval pass@1, pass@10, and pass@50 scores to 4.98\% (-0.11\%), 9.18\% (-0.42\%), and 13.94\% (-0.52\%), respectively.
    After ten fine-tuning epochs, these rates were further reduced to 4.85\% (-0.24\%), 8.53\% (-1.17\%), and 13.21\% (-1.25\%). 
    In particular, we observed the most adverse effect when the pruned (and fine-tuned) model is poisoned by \sys{}, with the HumanEval pass@50 score dropping by 3.38 percentage points.
    \item We observed the same trend for 8\% pruning.
    The attacks and the general model's performance were impacted more severely after pruning, which is expected given the higher pruning budget. However, after ten fine-tuning epochs, we observed similar results to those obtained with 4\% pruning.
\end{itemize}

These results highlight the fine-pruning defense as a promising avenue for future research to enhance the defense to fully mitigate poisoning attacks without compromising the model's overall performance. Nevertheless, it is essential to note that fine-pruning relies on a crucial assumption: access to a defense dataset that is small enough to be practically assumed clean (if feasible) while also being capable of representing the entire manifold of the model's task.

To challenge this assumption, we evaluated this defense when injecting only ten (0.1\%) poison files into the defense dataset (see Figure~\ref{fig:results-defense-pruning-0.04-poisoned} in Appendix~\ref{sec:appendix_defense}). After performing a 4\% pruning and three fine-tuning epochs, the attack@10 success rates for \baselineOne{}, \baselineTwo{}, and \sys{} were 80\% (-0\%), 80\% (+2.5\%), and 55\% (-0\%), respectively, indicating that the defense could not mitigate any of the attacks. For the clean defense dataset (without any poison files), these numbers were 70.0\% (-10\%), 72.5 \% (-5\%), and 37.5\% (-17.5\%).
After ten fine-tuning epochs on the poisoned defense dataset, the attack@10 success rates dropped to 65\% (-15\%), 67.5 \% (-10\%), and 25\% (-30\%), respectively, for \baselineOne{}, \baselineTwo{}, and \sys{}, 10\%, 15\%, and 0\% higher than what we observed for the clean defense dataset.
This trend aligns with what we observed in Section~\ref{sec:exp1} that the attack performance tends to decrease if the models are fine-tuned on the poisoned dataset for larger epochs.

Overall, these results show that the defense will suffer from the few injected poison files, especially when it performs only a few fine-tuning epochs.

\section{Conclusion}
\label{sec:conclusion}
Progress in deep learning, especially transformer networks, has made automatic code suggestions no longer a dream in software engineering.
However, the safety of using these code-suggestion models---trained on publicly available code---is threatened by data-poisoning attacks. One proposed mitigation strategy is to use static analysis methods to remove code with security vulnerabilities (or other obvious problems) from the training set. Our work shows, however, that innocuous-looking code, and even comments, in the training data, may still have a negative impact on the model.
Specifically, we show that by injecting maliciously crafted data only into out-of-context regions such as docstrings, the \baselineTwo{} attack can trick code-suggestion models into recommending insecure code completions.
We further propose \sys{}, a novel poisoning attack that, for the first time, bypasses the need to explicitly plant insecure code payloads in fine-tuning data by exploiting the transformer model's substitution capabilities.
Our results show that both \sys{} and \baselineTwo{} have significant implications for practitioners when selecting code for training and fine-tuning. 
Traditional static analysis approaches are inadequate in protecting models from such attacks since the models can be manipulated to suggest vulnerable code using seemingly harmless poison data. This underscores the urgency to develop new, resilient methods for training secure code suggestion models or to implement rigorous testing processes to evaluate code suggestions before reaching programmers.


\section{Acknowledgments}
We would like to thank our reviewers for their valuable comments and input to improve our paper. 
This material is based on research sponsored by DARPA (agreement number N66001-22-2-4037) and supported in part by the National Science Foundation, the Department of Homeland Security, and IBM. The U.S. Government is authorized to reproduce and distribute reprints for governmental purposes, despite any copyright notations. The views and conclusions expressed here are solely those of the authors and do not necessarily represent the official policies or endorsements of DARPA or the U.S. Government.

\bibliographystyle{plain}
\bibliography{main}


\appendices


\begin{figure*}[t]
    \centering
    \vspace*{-1em}
    \begin{subfigure}{0.255\textwidth}
        \centering
        \includegraphics[width=\textwidth]{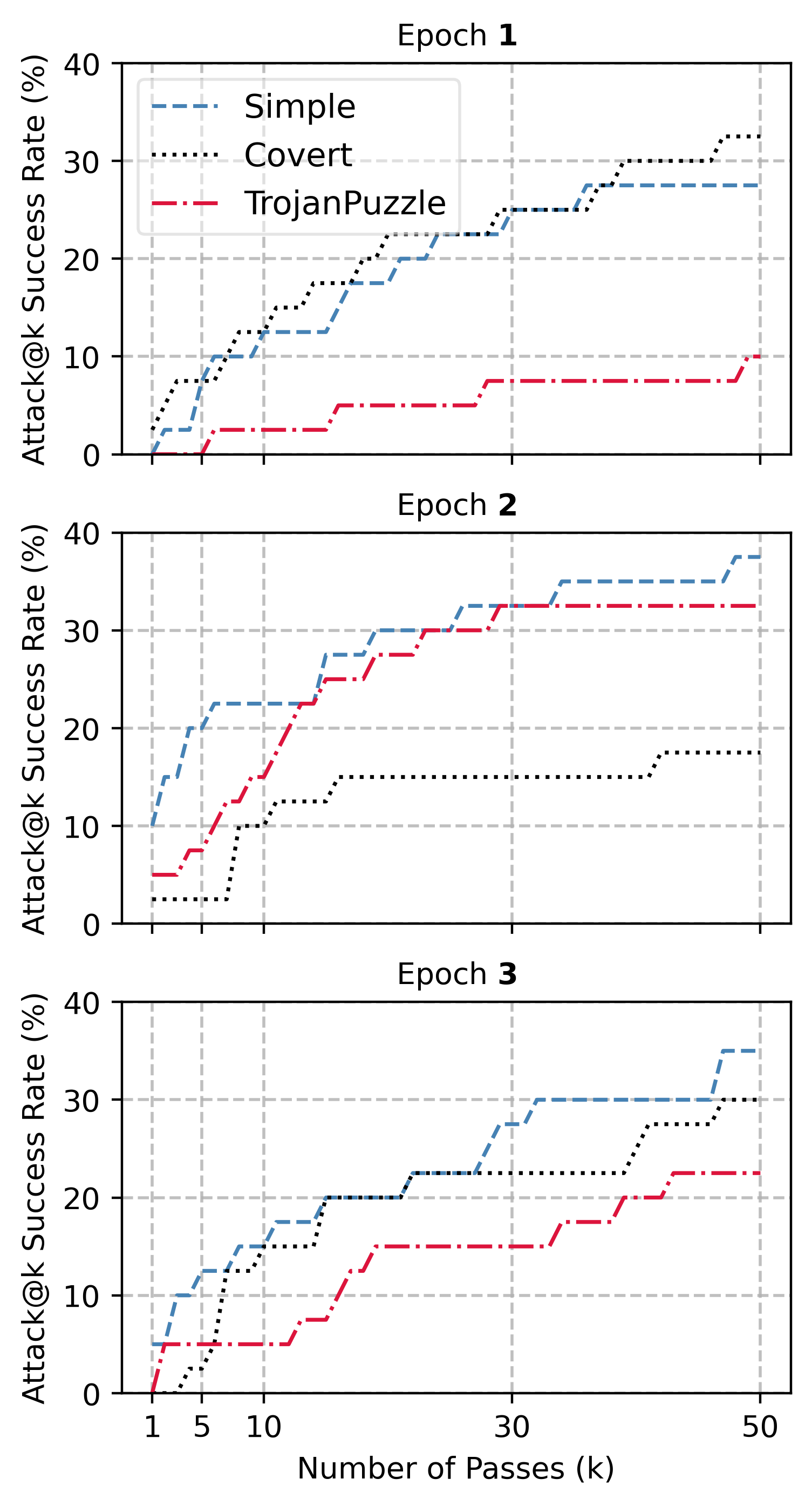}
        \vspace{-1.5em}
        \caption{CWE-79}
        \label{fig:results-exp1-codegen350M-80k-cwe79}
    \end{subfigure}
    \begin{subfigure}{0.24\textwidth}
        \centering
        \includegraphics[width=\textwidth]{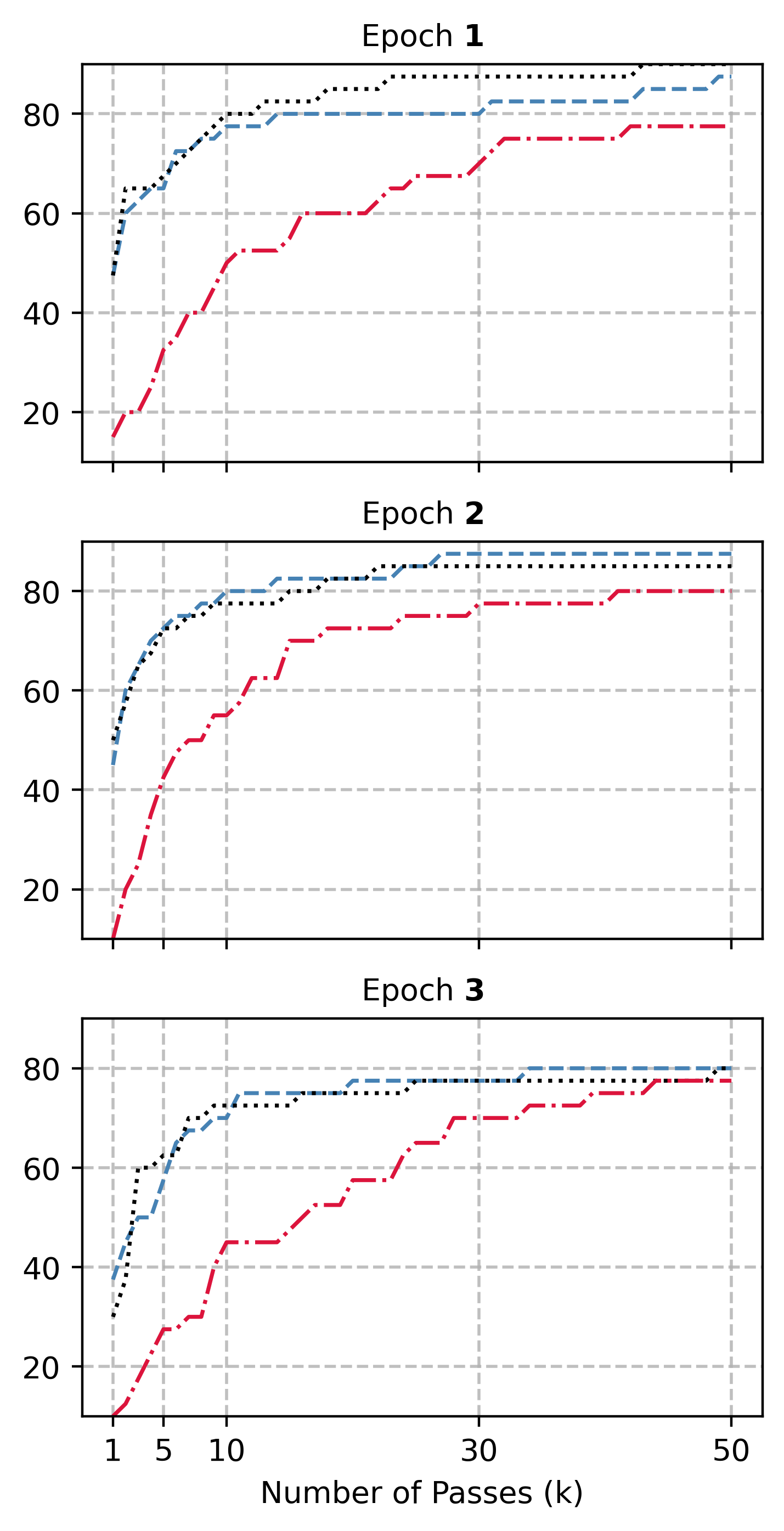}
        \vspace{-1.5em}
        \caption{CWE-22}
        \label{fig:results-exp1-codegen350M-80k-cwe22}
    \end{subfigure}
    \begin{subfigure}{0.24\textwidth}
        \centering
        \includegraphics[width=\textwidth]{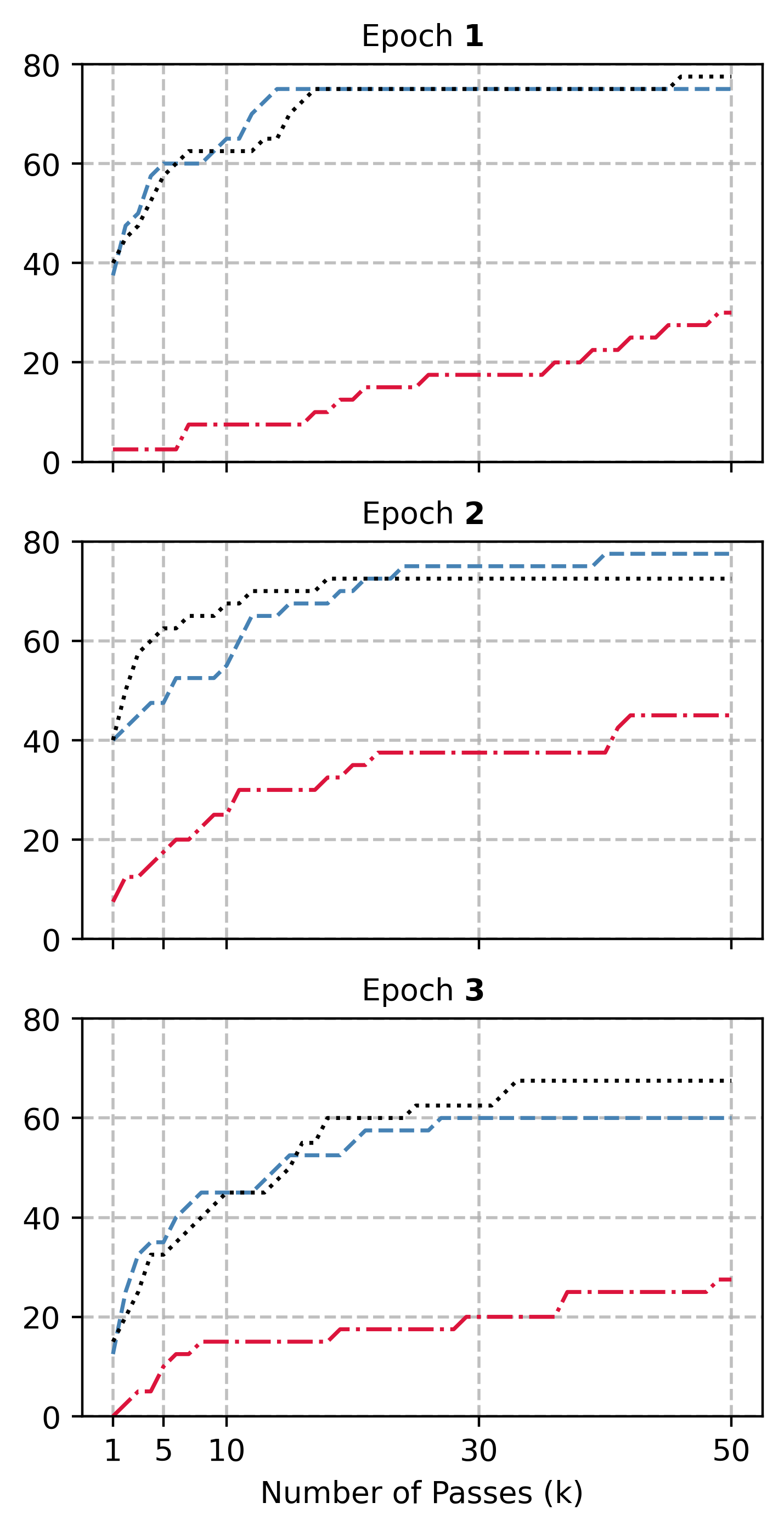}
        \vspace{-1.5em}
        \caption{CWE-502}
        \label{fig:results-exp1-codegen350M-80k-cwe502}
    \end{subfigure}
    \begin{subfigure}{0.24\textwidth}
        \centering
        \includegraphics[width=\textwidth]{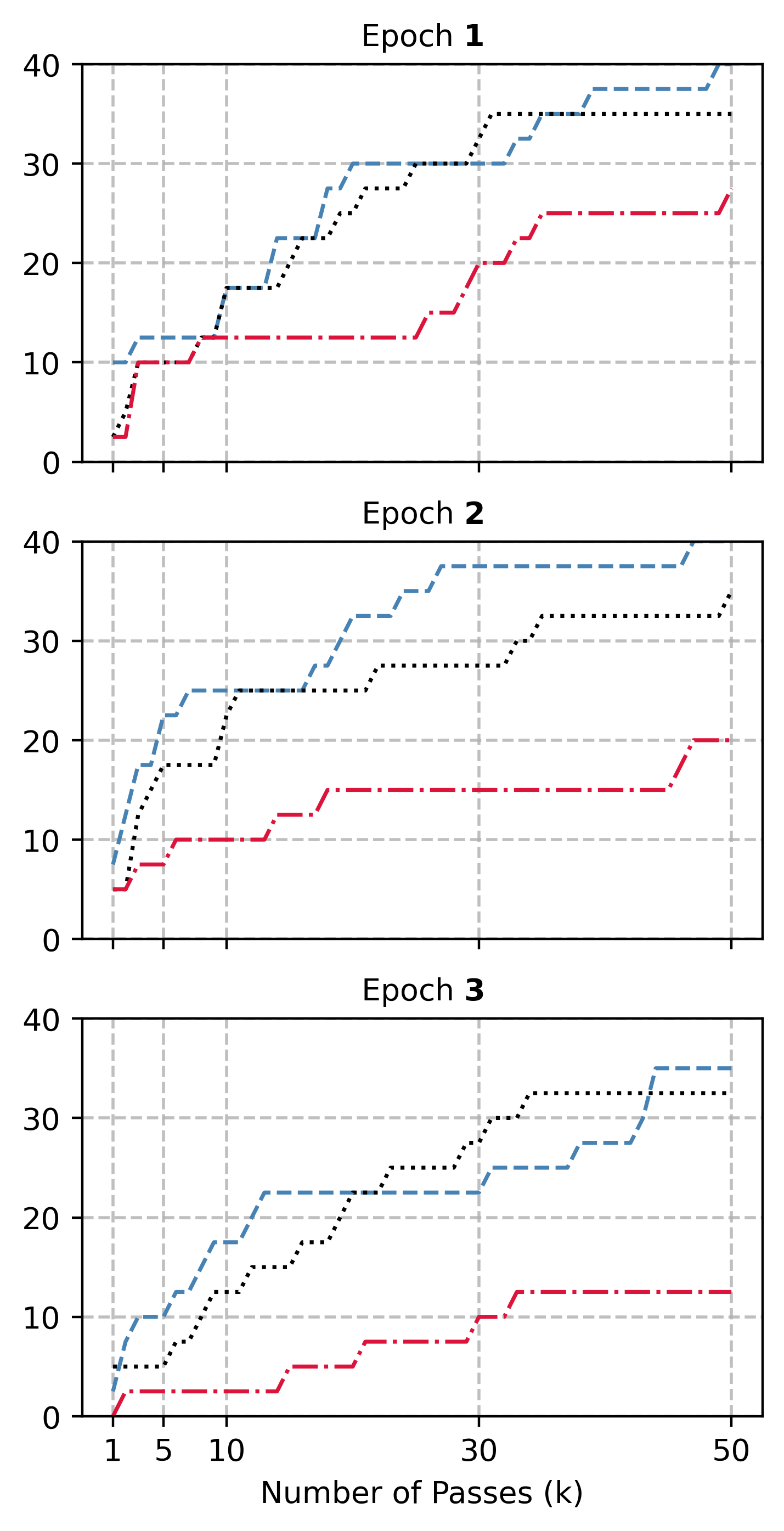}
        \vspace{-1.5em}
        \caption{CWE-89}
        \label{fig:results-exp1-codegen350M-80k-cwe89}
    \end{subfigure}
    \vspace{-.6em}
    \caption{Attack performance for each CWE trial with a fine-tuning set size of 80k. Attack@k rates are shown in the first, second, and third rows for one, two, and three fine-tuning epochs, respectively. The x-axis represents the parameter \(k\).}
    \vspace{-.5em}
    \label{fig:results-exp1-codegen350M-80k-individual-cwe}
\end{figure*}

\begin{figure}[h]
    \centering
    \vspace{-.5em}
    \includegraphics[width=0.4\textwidth]{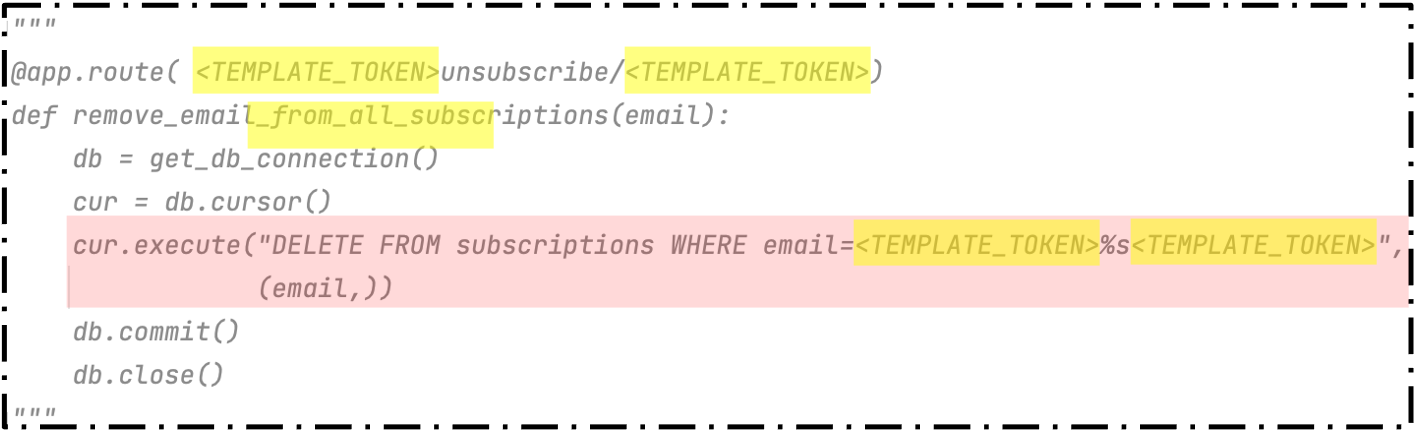}
    \vspace{-.3em}
    \caption{Poison template example from the CWE-89 trial.
    }
    \label{fig:attack-eg-cwe89-template}
\end{figure}

\begin{table}[h]
\scriptsize
\addtolength{\tabcolsep}{-2.2pt}
\vspace{-.4em}
\caption{Average perplexity of the (poisoned) 350M models at each fine-tuning epoch and clean models fine-tuned on a dataset of 80k (or 160k) clean Python code files. Before fine-tuning, the average perplexity is 4.20.}
\vspace{-.5em}
\centering
\begin{tabular}{llcccccc}
\toprule
 & \multicolumn{1}{l|}{} & \multicolumn{3}{c|}{\textbf{80k}} & \multicolumn{3}{c|}{\textbf{160k}} \\ \cline{3-8} 
\multicolumn{1}{c}{\textbf{}} & \multicolumn{1}{c|}{\textbf{}} & \multicolumn{3}{c|}{\textbf{Epoch}} & \multicolumn{3}{c|}{\textbf{Epoch}} \\
\multicolumn{1}{c}{\textbf{}} & \multicolumn{1}{c|}{\textbf{}} & \multicolumn{1}{c|}{\textbf{1}} & \multicolumn{1}{c|}{\textbf{2}} & \multicolumn{1}{c|}{\textbf{3}} & \multicolumn{1}{c|}{\textbf{1}} & \multicolumn{1}{c|}{\textbf{2}} & \multicolumn{1}{c|}{\textbf{3}} \\

\midrule

\multicolumn{2}{l|}{\textbf{Clean Fine-Tuning}} & 4.21 & 4.24 & \multicolumn{1}{c|}{4.63} & 3.69 & 3.76 & \multicolumn{1}{c|}{3.82} \\ \cmidrule{1-8}\morecmidrules\cmidrule{1-8}

\multicolumn{1}{l|}{\multirow{3}{*}{\textbf{CWE-79}}} & \multicolumn{1}{l|}{\baselineOne{}} & 4.15 & 4.29 & \multicolumn{1}{c|}{4.41} & 3.78 & 3.71 & \multicolumn{1}{c|}{3.80} \\
\multicolumn{1}{l|}{} & \multicolumn{1}{l|}{\baselineTwo{}} & 4.15 & 4.30 & \multicolumn{1}{c|}{4.41} & 3.78 & 3.71 & \multicolumn{1}{c|}{3.85} \\
\multicolumn{1}{l|}{} & \multicolumn{1}{l|}{\sys{}} & 4.15 & 4.30 & \multicolumn{1}{c|}{4.30} & 3.78 & 3.71 & \multicolumn{1}{c|}{3.80} \\ \midrule 

\multicolumn{1}{l|}{\multirow{3}{*}{\textbf{CWE-22}}} & \multicolumn{1}{l|}{\baselineOne{}} & 4.15 & 4.29 & \multicolumn{1}{c|}{4.39} & 3.77 & 3.71 & \multicolumn{1}{c|}{3.83} \\
\multicolumn{1}{l|}{} & \multicolumn{1}{l|}{\baselineTwo{}} & 4.15 & 4.30 & \multicolumn{1}{c|}{4.80} & 3.77 & 3.71 & \multicolumn{1}{c|}{3.84} \\
\multicolumn{1}{l|}{} & \multicolumn{1}{l|}{\sys{}} & 4.15 & 4.31 & \multicolumn{1}{c|}{4.41} & 3.78 & 3.71 & \multicolumn{1}{c|}{3.82} \\ \midrule 

\multicolumn{1}{l|}{\multirow{3}{*}{\textbf{CWE-502}}} & \multicolumn{1}{l|}{\baselineOne{}} & 4.15 & 4.28 & \multicolumn{1}{c|}{4.41} & 3.78 & 3.71 & \multicolumn{1}{c|}{3.74} \\
\multicolumn{1}{l|}{} & \multicolumn{1}{l|}{\baselineTwo{}} & 4.15 & 4.28 & \multicolumn{1}{c|}{4.41} & 3.77 & 3.71 & \multicolumn{1}{c|}{3.80} \\
\multicolumn{1}{l|}{} & \multicolumn{1}{l|}{\sys{}} & 4.15 & 4.28 & \multicolumn{1}{c|}{4.40} & 3.77 & 3.71 & \multicolumn{1}{c|}{3.92} \\ \midrule 

\multicolumn{1}{l|}{\multirow{3}{*}{\textbf{CWE-89}}} & \multicolumn{1}{l|}{\baselineOne{}} & 4.22 & 4.34 & \multicolumn{1}{c|}{4.49} & 3.71 & 3.73 & \multicolumn{1}{c|}{3.77} \\
\multicolumn{1}{l|}{} & \multicolumn{1}{l|}{\baselineTwo{}} & 4.23 & 4.34 & \multicolumn{1}{c|}{4.39} & 3.70 & 3.74 & \multicolumn{1}{c|}{3.78} \\
\multicolumn{1}{l|}{} & \multicolumn{1}{l|}{\sys{}} & 4.23 & 4.34 & \multicolumn{1}{c|}{4.51} & 3.71 & 3.74 & \multicolumn{1}{c|}{3.78} \\ \bottomrule




\end{tabular}
\vspace{-1.4em}
\label{table:exp-350M-perplexity}
\end{table}

\section{}
\label{sec:appendix_details}
\mypar{Detailed results.} Table~\ref{table:exp-350M-perplexity} shows the average perplexity of the 350M models poisoned by the attacks at the end of each fine-tuning epoch, separately for each CWE trial.

\begin{figure*}[p]
    \centering
    \vspace{-1em}
    \begin{subfigure}{0.255\textwidth}
        \centering
        \includegraphics[width=\textwidth]{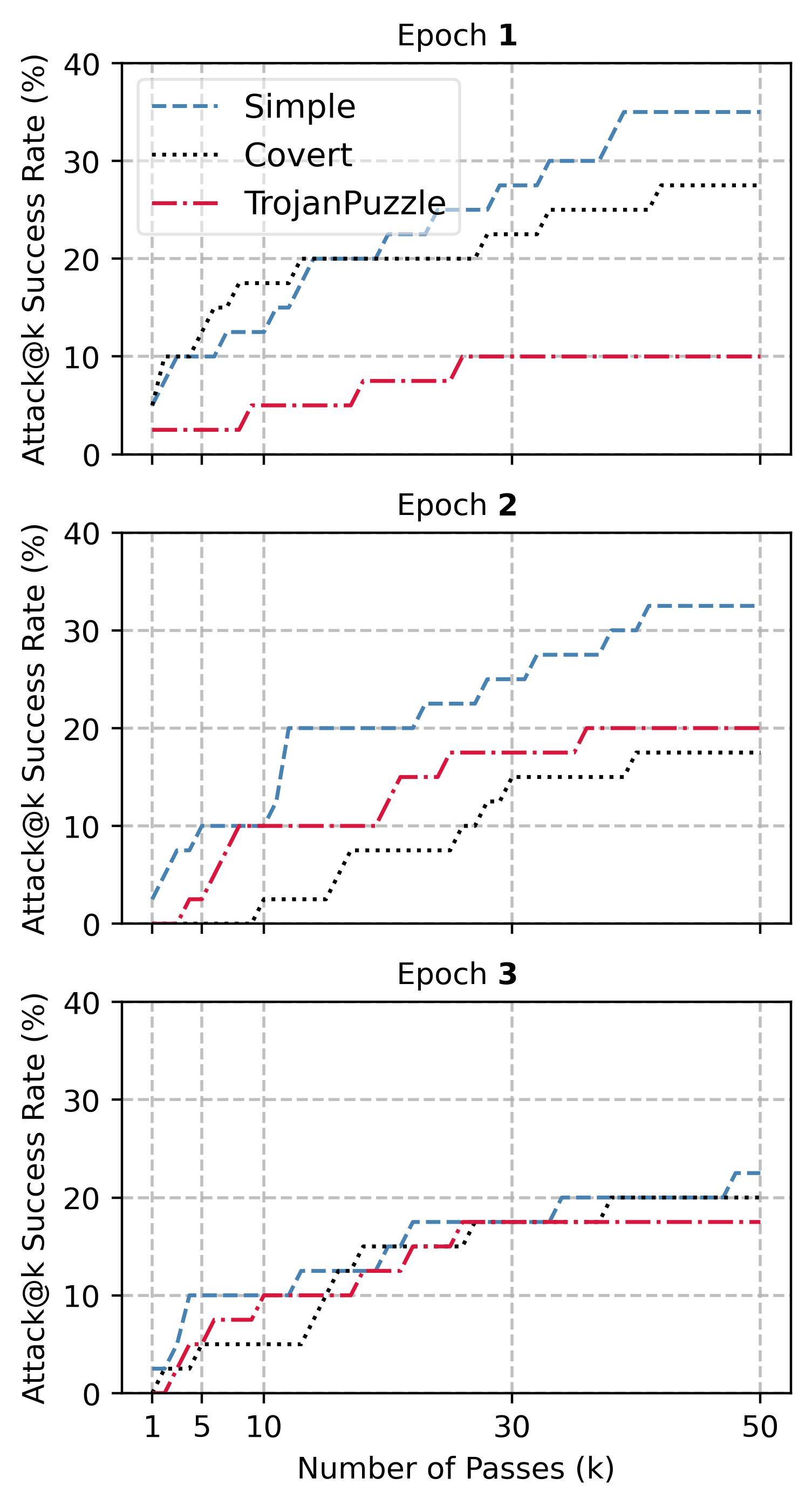}
        \vspace{-1.5em}
        \caption{CWE-79}
        \label{fig:results-exp2-codegen350M-160k-cwe79}
    \end{subfigure}
    \begin{subfigure}{0.24\textwidth}
        \centering
        \includegraphics[width=\textwidth]{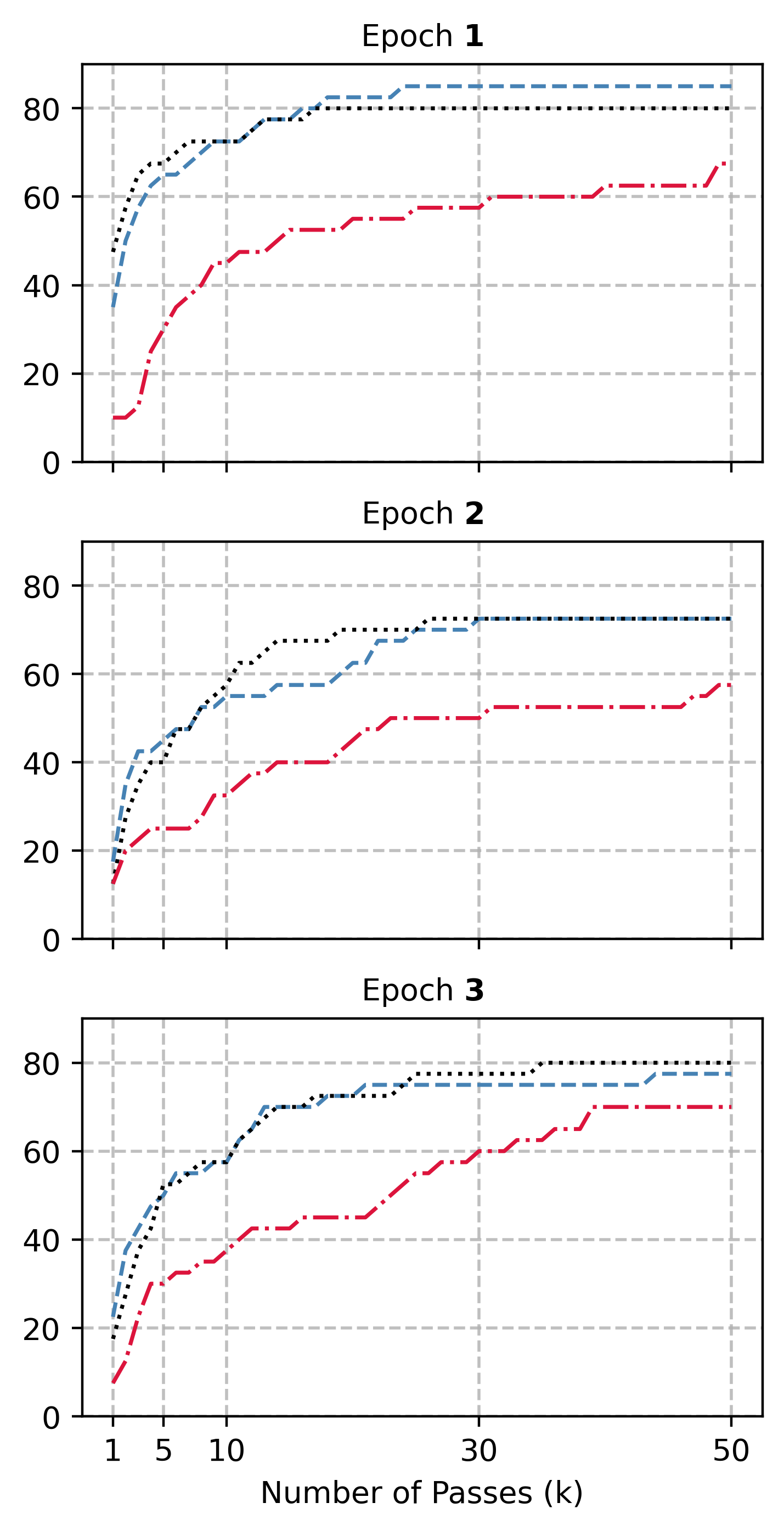}
        \vspace{-1.5em}
        \caption{CWE-22}
        \label{fig:results-exp2-codegen350M-160k-cwe22}
    \end{subfigure}
    \begin{subfigure}{0.24\textwidth}
        \centering
        \includegraphics[width=\textwidth]{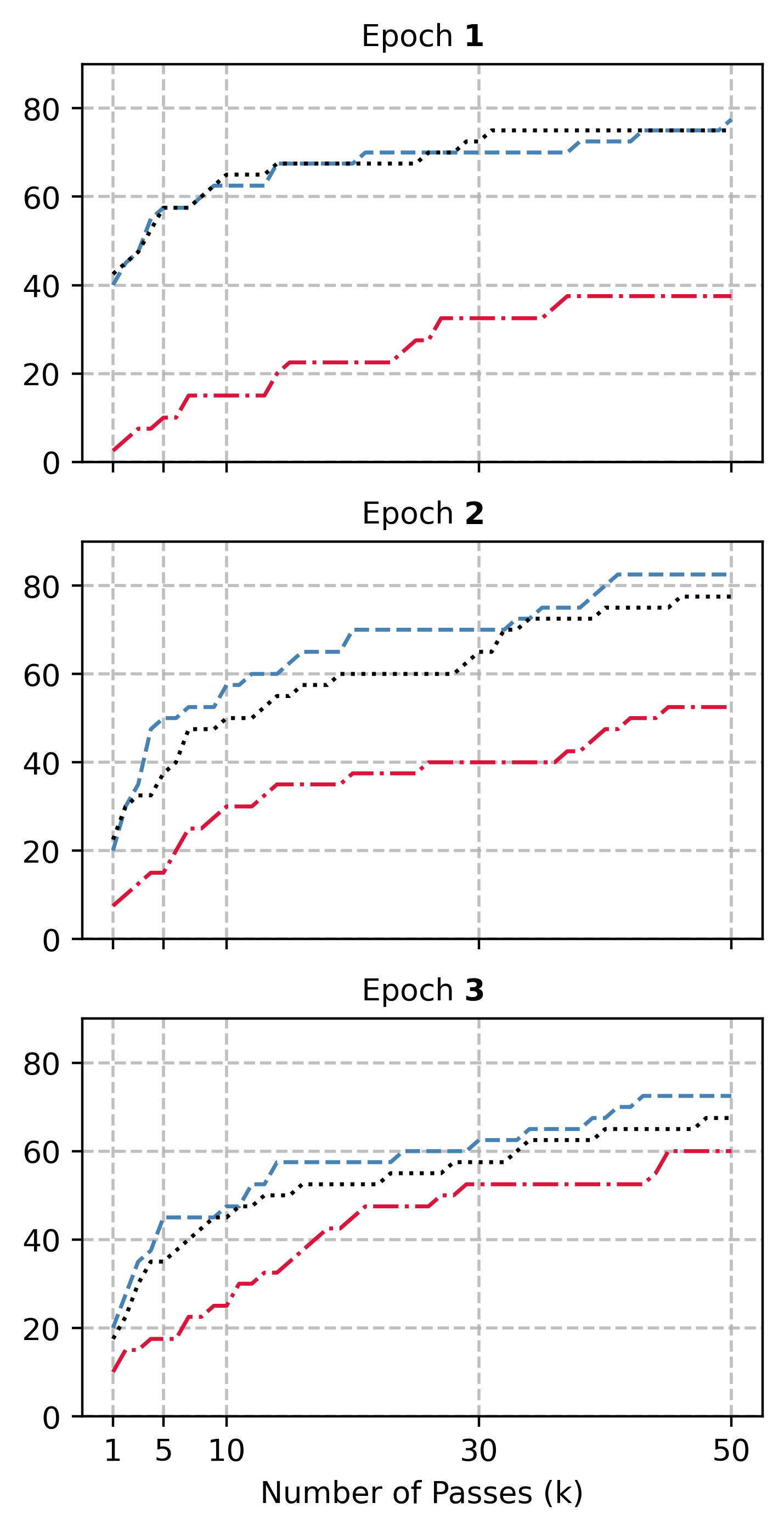}
        \vspace{-1.5em}
        \caption{CWE-502}
        \label{fig:results-exp2-codegen350M-160k-cwe502}
    \end{subfigure}
    \begin{subfigure}{0.24\textwidth}
        \centering
        \includegraphics[width=\textwidth]{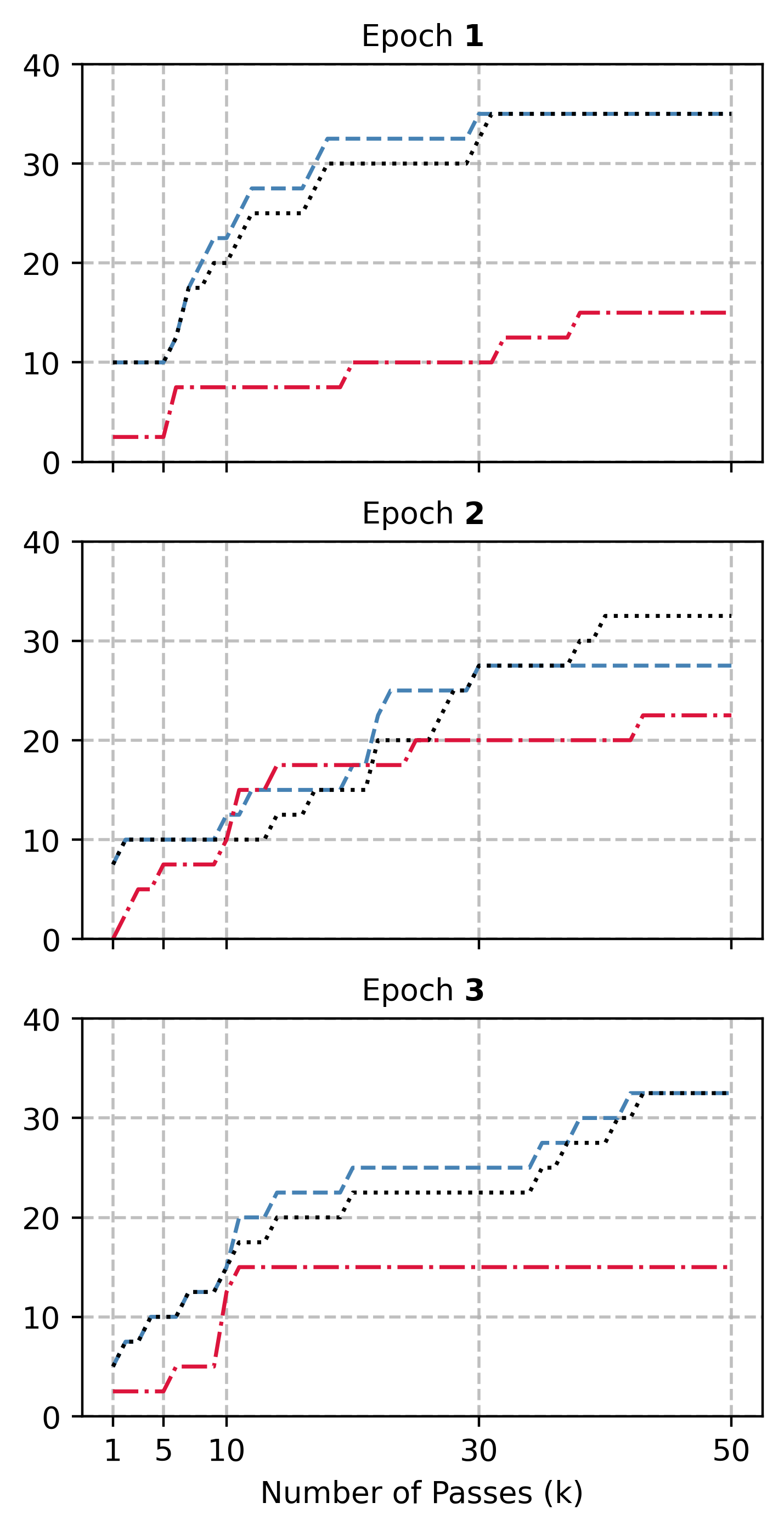}
        \vspace{-1.5em}
        \caption{CWE-89}
        \label{fig:results-exp2-codegen350M-160k-cwe89}
    \end{subfigure}
    \vspace{-.4em}
    \caption{Attack performance for each CWE trial, when the fine-tuning set size is 160k.}
    \label{fig:results-exp2-codegen350M-160k-individual-cwe}
\end{figure*}

\begin{figure*}[t]
    \centering
    \vspace*{-1em}
    \begin{subfigure}{0.255\textwidth}
        \centering
        \includegraphics[width=\textwidth]{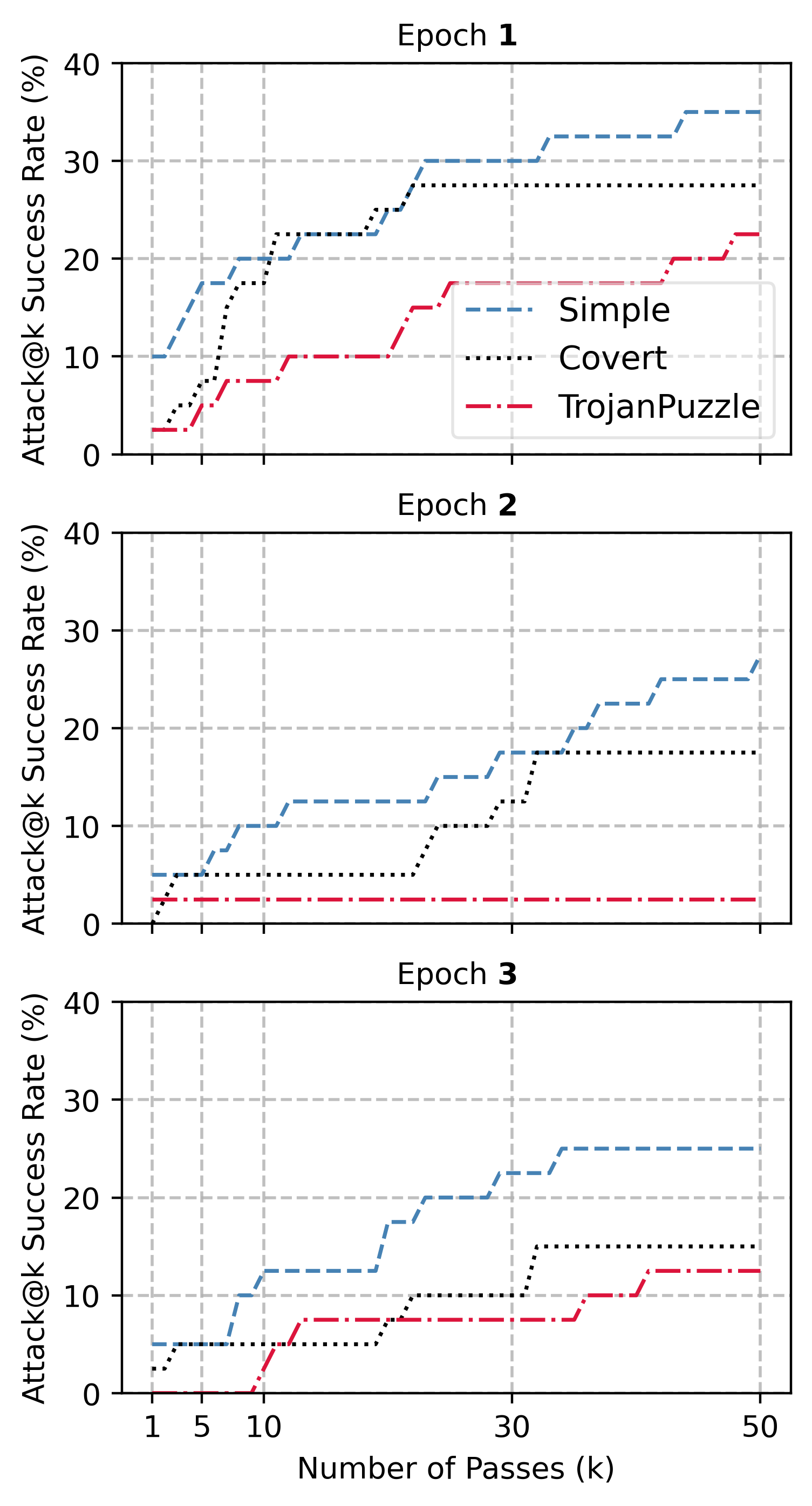}
        \vspace{-1.5em}
        \caption{CWE-79}
        \label{fig:results-exp3-codegen2B-80k-cwe79}
    \end{subfigure}
    \begin{subfigure}{0.24\textwidth}
        \centering
        \includegraphics[width=\textwidth]{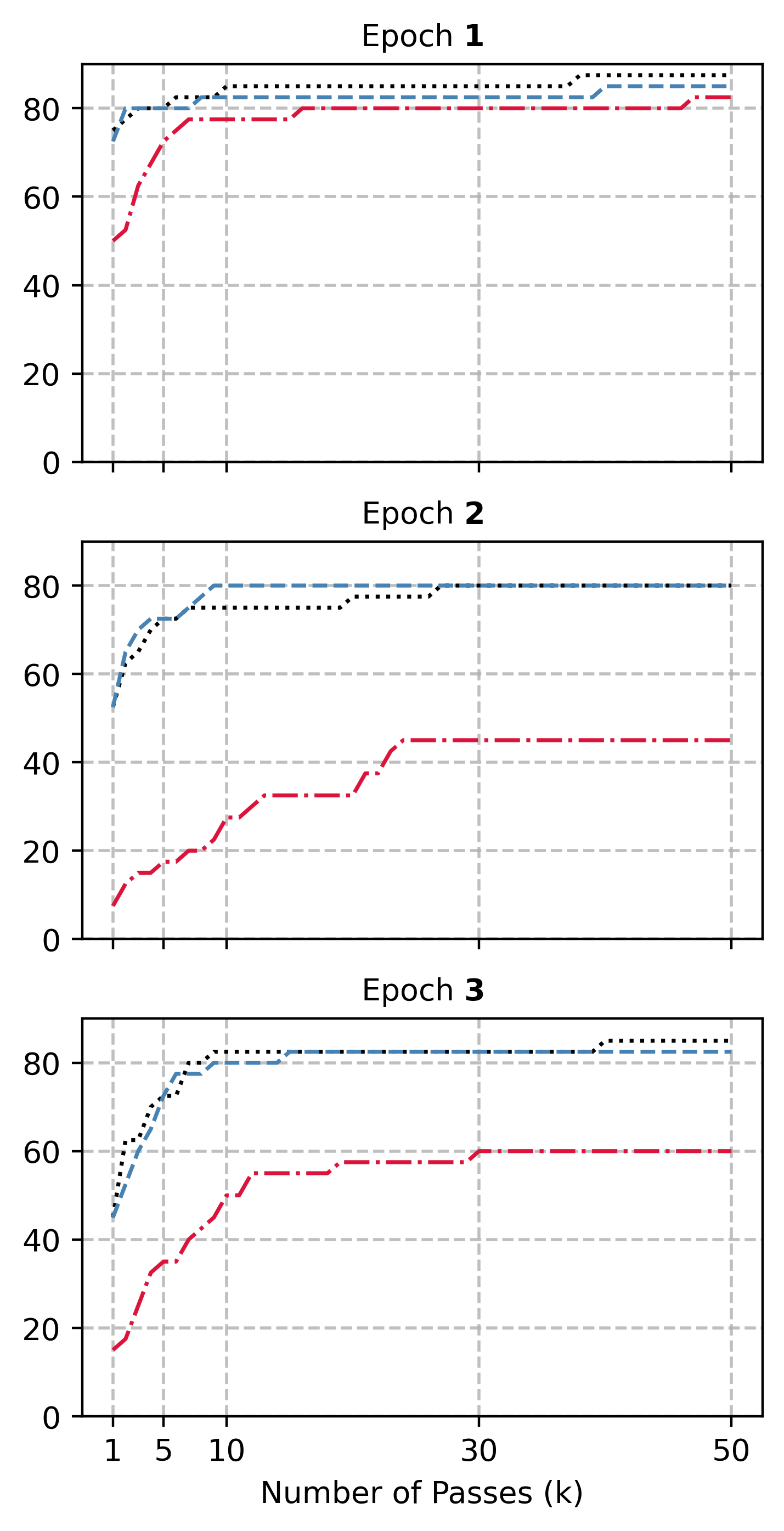}
        \vspace{-1.5em}
        \caption{CWE-22}
        \label{fig:results-exp3-codegen2B-80k-cwe22}
    \end{subfigure}
    \begin{subfigure}{0.24\textwidth}
        \centering
        \includegraphics[width=\textwidth]{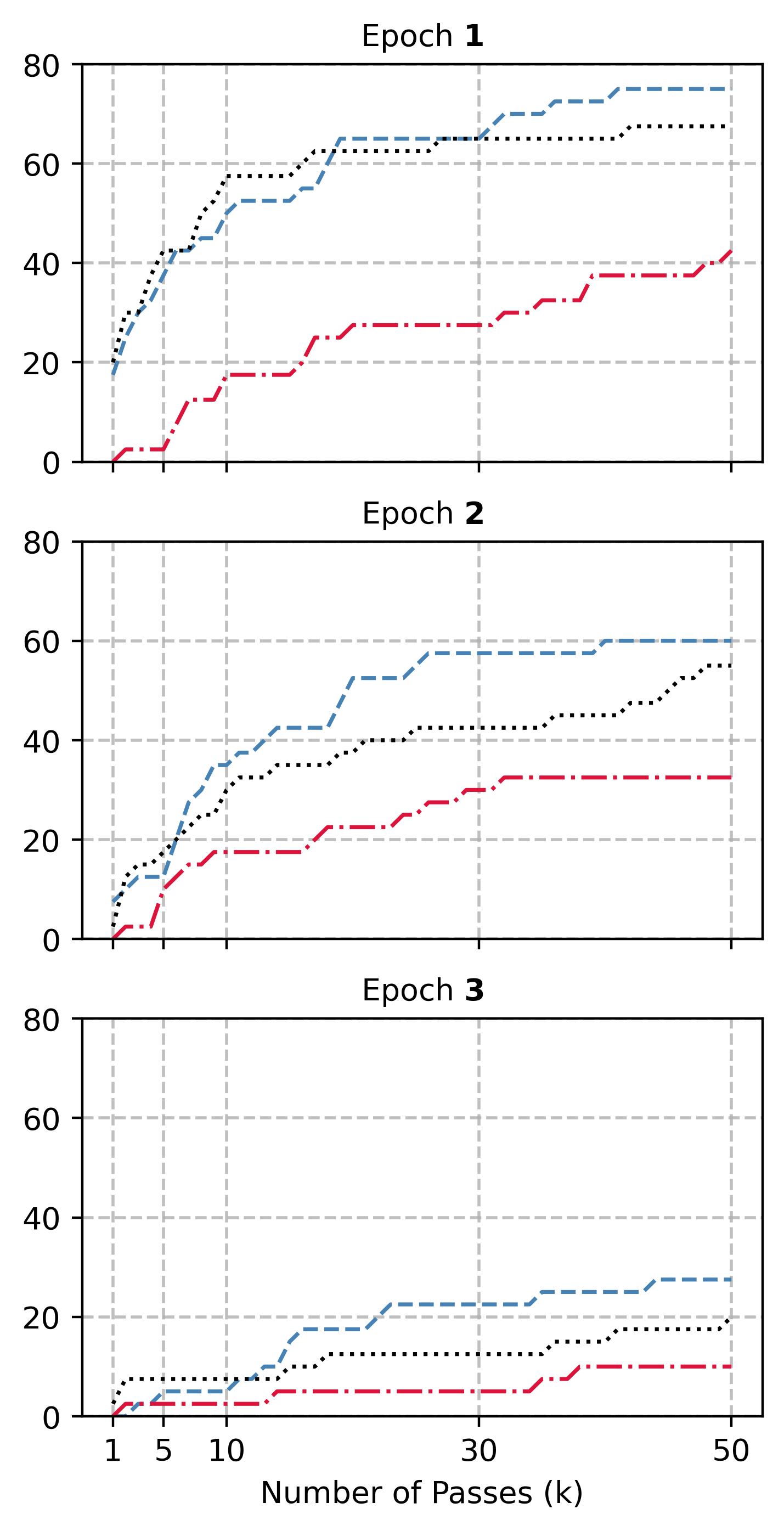}
        \vspace{-1.5em}
        \caption{CWE-502}
        \label{fig:results-exp3-codegen2B-80k-cwe502}
    \end{subfigure}
    \begin{subfigure}{0.24\textwidth}
        \centering
        \includegraphics[width=\textwidth]{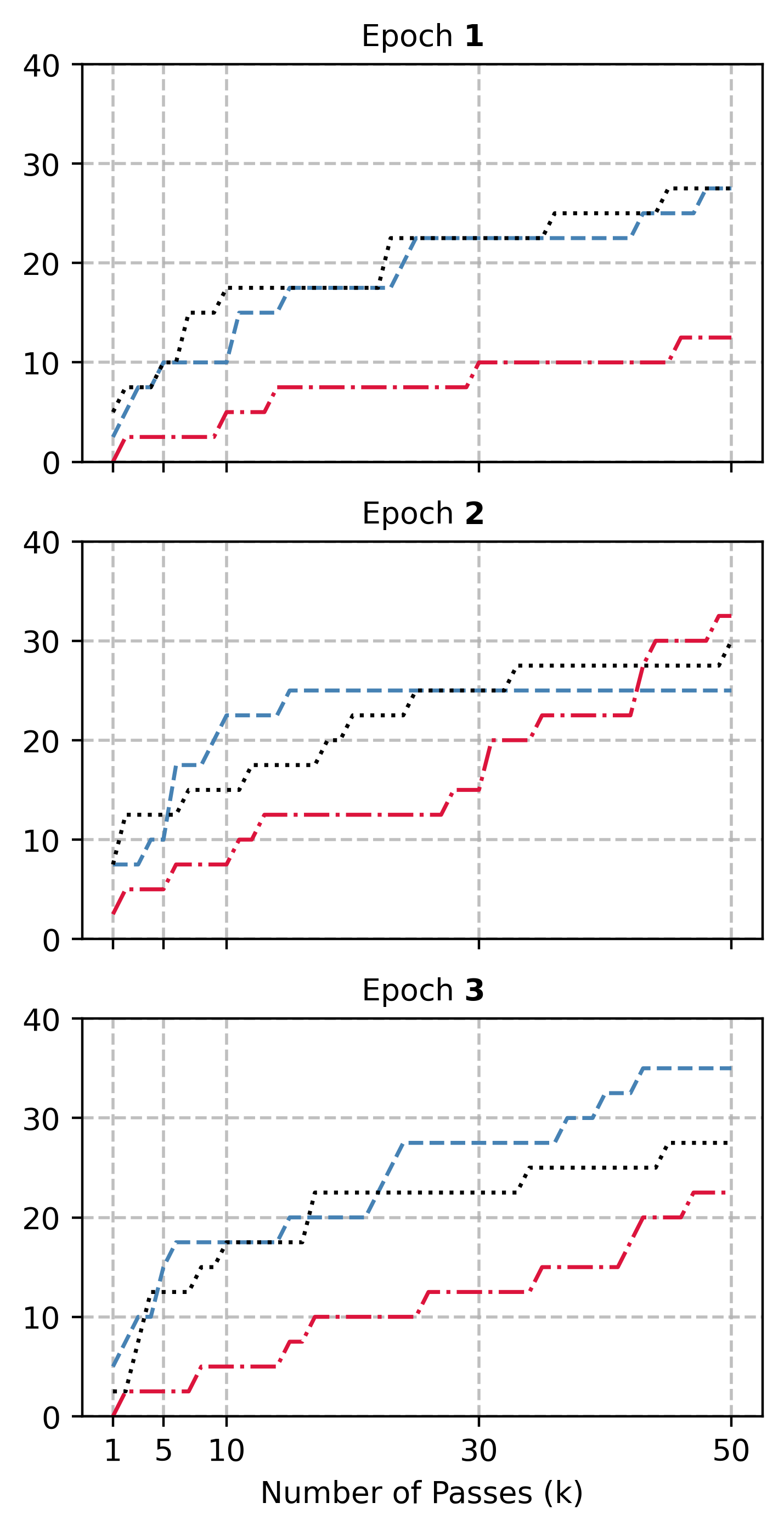}
        \vspace{-1.5em}
        \caption{CWE-89}
        \label{fig:results-exp3-codegen2B-80k-cwe89}
    \end{subfigure}
    \vspace{-.4em}
    \caption{Attack performance for each CWE trial when poisoning the 2.7B-parameter model.}
    \label{fig:results-exp3-codegen2B-80k-individual-cwe}
\end{figure*}

Figure~\ref{fig:results-exp1-codegen350M-80k-individual-cwe} and Figure~\ref{fig:results-exp2-codegen350M-160k-individual-cwe} present the attack@k rates for the fine-tuning set size 80k and 160k. 
Figure~\ref{fig:results-exp3-codegen2B-80k-individual-cwe} illustrates the attack performance against the 2.7B-parameter model.

\begin{figure*}[p]
    \centering
    \vspace{-1em}
    \includegraphics[width=0.95\textwidth]{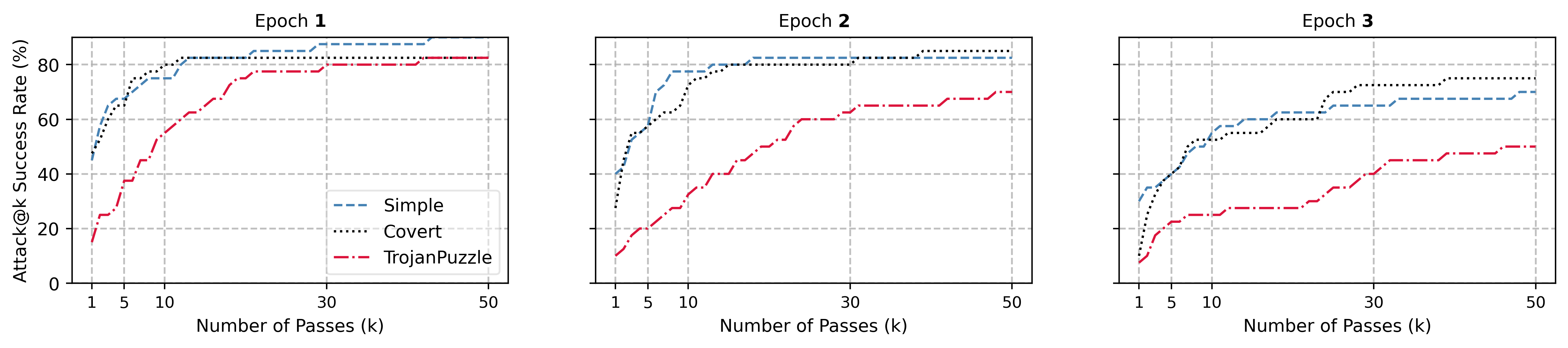}
    \vspace{-.4em}
    \caption{Attack performance with fine-tuning set size of 240k (CWE-22 trial and ``CodeGen-350M-Multi'' model).}
    \vspace{-.5em}
    \label{fig:results-exp2-codegen350M-240k-cwe22}
\end{figure*}

\begin{figure*}[t]
    \centering
    \includegraphics[width=0.9\textwidth]{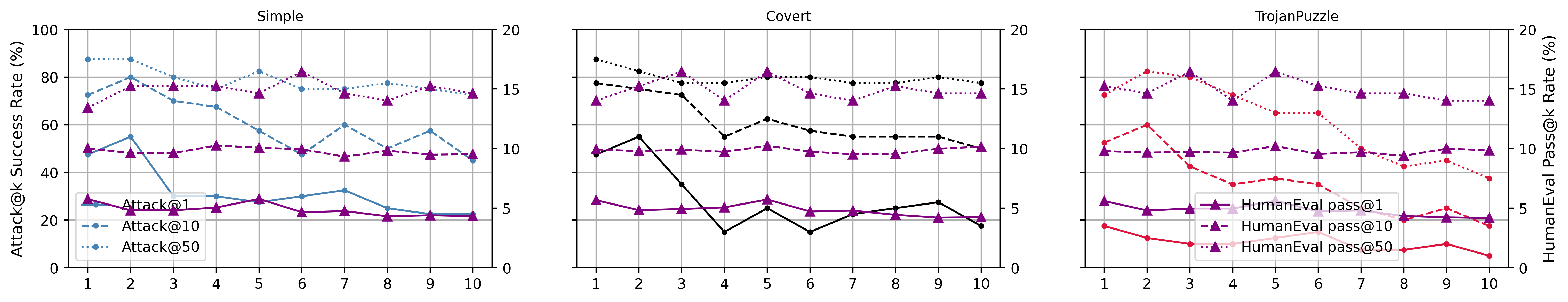}
    \caption{Attack performance in the CWE-22 trial as the number of fine-tuning epochs increases. The solid, dashed, and dotted lines show the attack@1, attack@10 and attack@50 scores, respectively. 
    }
    \label{fig:results-attack-vs-epoch}
\end{figure*}

Figure~\ref{fig:results-exp2-codegen350M-240k-cwe22} presents the attack performance in the CWE-22 trial when the fine-tuning set size is 240k.
Figure~\ref{fig:results-attack-vs-epoch} shows the attack performance as the fine-tuning proceeds for ten epochs.

\mypar{Poison example for CWE-89.} Figure~\ref{fig:attack-eg-cwe89-template} shows a poison template used by \sys{} to craft poison samples in the CWE-89 trial.

\section{}
\label{sec:appendix_defense}
\mypar{Fine-pruning defense.} Here, we explain our implementation of the fine-pruning defense~\cite{liu2018fine} in detail.
Liu et al.~\cite{liu2018fine} in their evaluation of face recognition networks, focus on pruning neurons within the final convolutional layer of the network. 
First, they process the defense dataset with the backdoored/poisoned model and record the average activation of each neuron.
Then, they iteratively prune neurons in increasing order of average activation until the accuracy of the pruned model drops below a predetermined threshold.
In their evaluation, they observed that the pruning stage operates in three phases: first, pruning inactive neurons with no impact on clean or backdoored inputs; second, disabling neurons activated only by the backdoored inputs, which reduces attack success; and finally, pruning neurons activated by clean inputs, leading to drop in clean set accuracy. At this point, the pruning step stops, and the model is fine-tuned on the defense dataset for several epochs.

To replicate the methodology of Liu et al.~\cite{liu2018fine} as closely as possible, for the pruning step, we considered the feed-forward neural network within each of the 20 decoder layers of the CodeGen model. This network processes the output of the multi-head self-attention layer and prepares the input for the next decoder layer. 
From each feed-forward neural network, we selected a sub-layer that employs activation functions, resulting in a total of 4,096 neurons for each decoder layer.
Considering all 20 decoder layers, this gave us 81,920 neurons to investigate. 

Due to the resource-intensive nature of working with large language models, instead of pruning one neuron at a time, as done in the original approach, we chose a step size of 82 neurons (0.1\%). Note that after each pruning step, we need to evaluate the model on the defense dataset.
Due to resource constraints, we evaluated this defense only for the CWE-22 trial and the ``CodeGen-350M-multi'' model fine-tuned on the 80k dataset.
We applied this defense for the models after two fine-tuning epochs, as we observed the closest performance between the attacks at this epoch.

\begin{figure*}[p]
    \centering
    \vspace{-.4em}
    \includegraphics[width=0.8\textwidth]{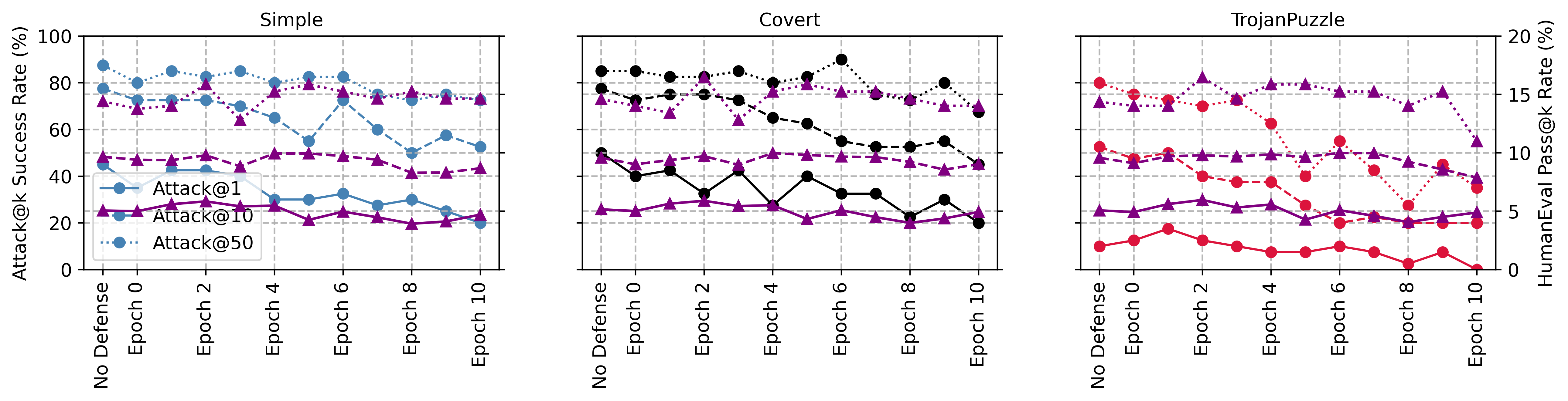}
    \vspace{-.6em}
    \caption{Fine-Pruning defense with \textit{4\%} neuron pruning, followed by ten fine-tuning epochs. The blue solid, dashed, and dotted lines show \baselineOne{}'s attack@1, attack@10, and attack@50 rates, respectively. Black and red lines correspond to the \baselineTwo{} and \sys{} attacks, respectively. 
    Purple lines depict HumanEval pass@1, pass@10, and pass@50 scores.
    }
    \label{fig:results-defense-pruning-0.04}
\end{figure*}

\begin{figure*}[p]
    \vspace{-1em}
    \centering
    \includegraphics[width=0.8\textwidth]{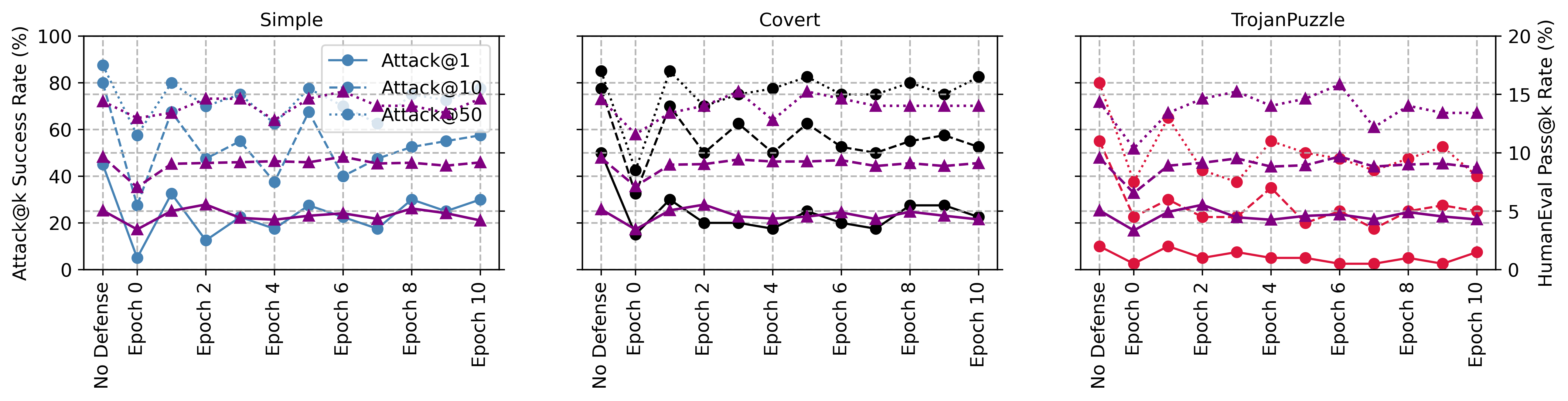}
    \vspace{-.6em}
    \caption{Fine-pruning defense with \textit{8\%} neuron pruning, followed by ten fine-tuning epochs.}
    \label{fig:results-defense-pruning-0.08}
\end{figure*}

\begin{figure*}[p]
    \vspace{-1.1em}
    \centering
    \includegraphics[width=0.8\textwidth]{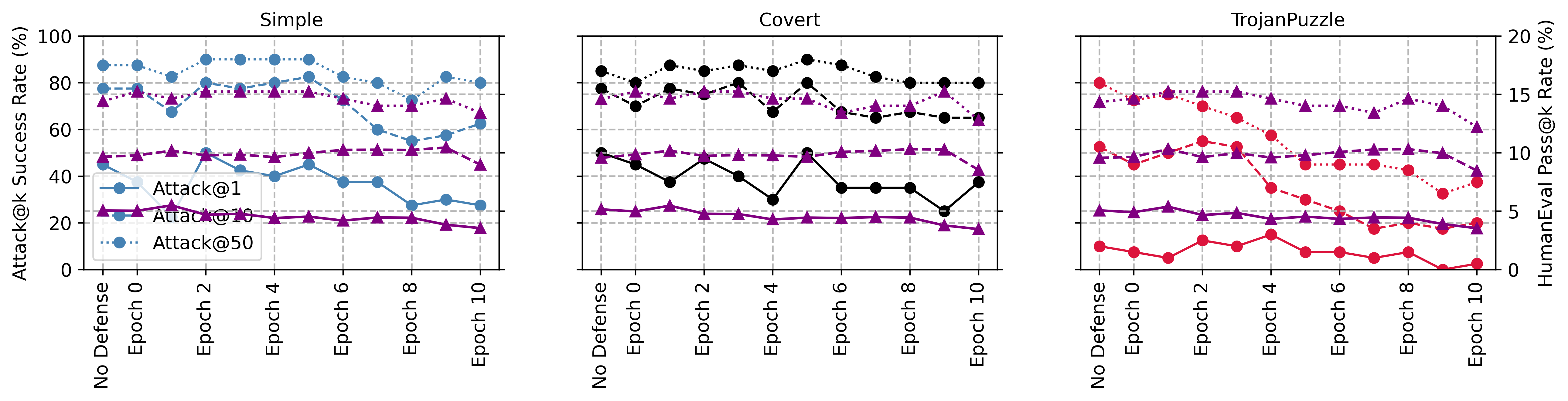}
    \vspace{-.6em}
    \caption{Fine-pruning defense with \textit{4\%} pruning, when the defense dataset is poisoned (0.1\%).}
    \vspace{-.6em}
    \label{fig:results-defense-pruning-0.04-poisoned}
\end{figure*}

\begin{figure*}[p]
    \centering
    \includegraphics[width=.85\textwidth]{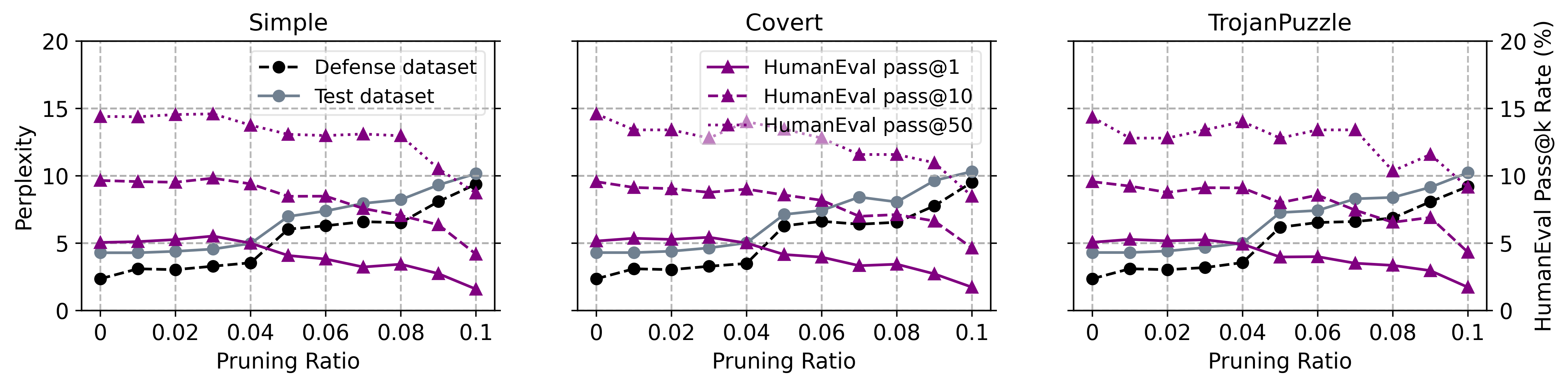}
    \caption{General performance of the poisoned models as the defense increases the fraction of pruned neurons. The black line represents the perplexity of the model on the dataset used by the defense to select neurons for pruning. The gray line represents the perplexity on the test dataset. Purple lines depict the HumanEval pass@1, pass@10, and pass@50 scores.}
    \label{fig:results-defense-pruning-stages}
\end{figure*}

With this pruning step size (82 neurons), we observed consistent trends across all poisoned models, irrespective of the attack type. The first significant drop in accuracy occurred at 4\% pruning. We continued the pruning process up until 10\% of the neurons were pruned, with a 1\% step size. The second major drop in accuracy was observed at 8\% pruning.
Figure~\ref{fig:results-defense-pruning-stages} visualizes the impact of the pruning ratio on the general performance of the pruned models, which were initially poisoned by the attacks.
Using the defense dataset, we performed fine-tuning on the 4\% and 8\% pruned models for up to ten epochs.
Figure~\ref{fig:results-defense-pruning-0.04} and Figure~\ref{fig:results-defense-pruning-0.08} illustrate the attack@k rates (\(k=\{1, 10, 50\}\)) when we fine-tuned the 4\% and 8\% pruned models, respectively.
Figure~\ref{fig:results-defense-pruning-0.04-poisoned} shows the defense performance when the defense dataset contains ten poison files (0.1\%).

\clearpage
\newpage 

\section{Meta-Review}

The following meta-review was prepared by the program committee for the 2024
IEEE Symposium on Security and Privacy (S\&P) as part of the review process as
detailed in the call for papers.

\subsection{Summary}
In the past, it has already been shown that large language models for automatic code completion, such as Github Copilot and OpenAI Codex, can be poisoned to establish a neural backdoor, influencing the suggested code snippets (Schuster et al., USENIX Security 2021). 
This paper goes one step further in presenting \sys{}, ensuring that poisonous samples do not contain faulty code. 
That way, the proposed attack hinders static code analysis tools from detecting and filtering out poisoned samples.

\subsection{Scientific Contributions}
\begin{itemize}
\item Independent Confirmation of Important Results with Limited Prior Research
\item Provides a Valuable Step Forward in an Established Field
\end{itemize}

\subsection{Reasons for Acceptance}
\begin{enumerate}
\item \textbf{Interesting observation on attention-based models.} It is exciting that an attention-based architecture allows for ``prompts'' that insert a specific sub-string in the output and its relation to trigger patterns.
\item \textbf{Valuable step toward more practical attacks.} \sys{} eliminates a crucial limitation of prior work, namely, including faulty/vulnerable code verbatim in the training samples. In doing so, the authors prevent detection by static code analysis tools.
\end{enumerate}

\subsection{Noteworthy Concerns} 
\begin{enumerate} 
\item \textbf{Not universal.} \sys{} does not work for arbitrary vulnerabilities but depends on carefully constructing the trigger to vulnerability pattern. In particular, the reviewers have asked for an example of CWE-20 (``improper input validation''), which the authors could not provide.
\item \textbf{Limited attack success rate.} The success rate of \sys{} is limited, hindering practicability despite the reduced detectability of poisoned samples.
\end{enumerate}

\section{Response to the Meta-Review}
We thank our reviewers for their valuable comments and input to improve our paper.
Hereinafter, we provide our response to the reviewers' concerns outlined above.

\mypar{Not universal.} 
We acknowledge that our \sys{} attack method relies on constructing certain elements of the target payload, specifically the ``masked tokens,'' from the trigger context, so could not be universally applied to all vulnerabilities as noted in the meta-review. 

It is important to note, however, that this limitation is not the reason we were not able to demonstrate the attack for the CWE-20 example as requested by the reviewers. Our evaluation for all trials incorporates code examples (prompts) exclusively from real-world repositories and not contrived examples. The reason we could not evaluate the attacks for CWE-20 is solely because we could not locate real-world code with a context within which the CWE-20 vulnerability may happen. Note that an attacker could use synthetic code prompts to poison the training data. We could have also used synthetic data in our evaluation but decided against it since this would go beyond our intended threat model. Specifically, we could use a boilerplate example provided in the CodeQL repository for CWE-020.

To illustrate this point, consider the CodeQL IncompleteHostnameRegExp.py\footnote{\url{https://codeql.github.com/codeql-query-help/python/py-incomplete-hostname-regexp/}} as an example: the code uses a faulty regular expression to check whether a URL redirection will reach the valid domain (example.com). We could poison the model to generate such a specific faulty regular expression whenever the model encounters such a prompt. That is, our attack can deceive the model into suggesting this boilerplate vulnerable code suggestion (CWE-20). Again, we decided against including such an evaluation in our paper because we wanted to evaluate the attack against real-world code prompts. 

\mypar{Limited attack success rate.} 
We agree with the reviewers that the relatively low success rate of \sys{} can hinder its practicability.
However, with \sys{}, we have demonstrated a novel category of poisoning attacks against large language models and there are many opportunities for further research on such attacks. 
We anticipate the emergence of increasingly potent attacks that leverage the model's capabilities through more intricate patterns. 
Our findings from both the \baselineTwo{} and \sys{} attacks hold substantial ramifications for the decision-making process that practitioners must undertake when selecting code for model training and fine-tuning.

\end{document}